\documentclass[10pt]{article}
\usepackage{cite}
\usepackage{graphicx}
\usepackage[english]{babel}
\usepackage{amsmath}
\usepackage{color}
\usepackage{caption}
\usepackage{cite}
\usepackage{hyperref}
\usepackage[toc,page]{appendix}
\usepackage{amssymb}
\usepackage{epstopdf}
\makeatletter
\newcommand*{\rom}[1]{\expandafter\@slowromancap\romannumeral #1@}
\makeatother

\begin{document}
\begin{center}
{\Large\bf Study of Electroweak Vacuum Stability from Extended Higgs Portal
of Dark Matter and Neutrinos}
\\
\vskip .5cm
{Purusottam Ghosh$^{a,}$\footnote{pghoshiitg@gmail.com},
Abhijit Kumar Saha$^{a,}$\footnote{abhijit.saha@iitg.ernet.in},
Arunansu Sil$^{a,}$\footnote{asil@iitg.ernet.in}}\\[3mm]
{\it{
$^a$ Department of Physics, Indian Institute of Technology Guwahati, 781039 Assam, India}
}
\end{center}

\vskip 1cm

\begin{abstract}

\noindent We investigate the electroweak vacuum stability in an extended version
of the Standard Model which incorporates two additional singlet scalar fields and three right handed neutrinos.
One of these extra scalars plays the role of dark matter while the other scalar not only helps in making the electroweak vacuum 
stable but also opens up the low mass window of the scalar singlet dark matter ($<$ 500 GeV). We consider the effect of large 
neutrino Yukawa coupling on 
the running of Higgs quartic coupling. We have analyzed the constraints on 
the model and identify the range of parameter space which is consistent with neutrino mass, appropriate relic density and direct 
search limits from the latest XENON 1T preliminary result as well as  realizing the 
  stability of the electroweak vacuum upto the Planck scale. 
\end{abstract}

\section{Introduction}
The discovery of the Higgs boson\cite{Aad:2012tfa,Chatrchyan:2012xdj,Giardino:2013bma} has been considered as the greatest triumph in present day particle physics.
Although the experimental search is still on in order to investigate the Higgs boson's properties, several theoretical
and phenomenological reasons are there to push us toward hunting for an enlarged Higgs sector compared to the one present
in Standard model (SM). For example 
the Higgs quartic coupling $\lambda_H$ in SM becomes negative at a high 
energy scale ($\Lambda_I^{\textrm{SM}}\sim10^{10}$ GeV) leading to a possible instability of Higgs vacuum.
The present measured
values of Higgs mass$\sim 125.09$ GeV\cite{Agashe:2014kda} and top mass $\sim 173.2$ GeV\cite{Agashe:2014kda}, suggest that 
the electroweak (EW) vacuum can be metastable
\cite{Isidori:2001bm, Buttazzo:2013uya, Anchordoqui:2012fq, Tang:2013bz ,Degrassi:2012ry, Ellis:2009tp, EliasMiro:2011aa}. 
 However the conclusion exclusively depends on the precise measurement of the top mass\cite{Branchina:2014usa,Degrassi:2014hoa}. 
Also the metastability of the Universe
 is not a very robust situation in the context of cosmological inflation\cite{Kobakhidze:2013tn}. One of the possible solutions to this is to introduce new 
physics between EW scale and $\Lambda_I^{\textrm{SM}}$. In view of SM's incompetence in resolving some of the
other issues like dark matter, neutrino mass, matter antimatter symmetry, inflation etc, the introduction of new physics is of course a welcome feature.

In particular, SM fails to accommodate a significant share in terms of its content called dark matter(DM).
The most economical and popular scenario is the singlet scalar extension of the SM\cite{Silveira:1985rk,McDonald:1993ex,
Cline:2013gha,Guo:2010hq,Queiroz:2014yna,Ghorbani:2014gka,Bhattacharya:2016qsg,Bhattacharya:2016ysw,Casas:2017jjg} having Higgs portal interaction.
The stability of the dark matter is ensured by imposing a $Z_2$ symmetry on it.
The relic abundance and corresponding direct detection cross section are solely determined by the
DM (the scalar singlet)  mass and its coupling with Higgs ( portal coupling). However present experiments, LUX\cite{Akerib:2016vxi} and XENON1T\cite{Aprile:2017iyp},
strongly disfavor the model below $m_{\textrm{DM}}<500$ GeV except the resonance region. The bound on
Higgs invisible decay width further constrain the model for $m_{\textrm{DM}}<62.5$ GeV\cite{Athron:2017kgt}. Hence
a large range of DM mass seems to be excluded within this simplest framework which otherwise would be an interesting
region of search for several ongoing and future direct\cite{Akerib:2016vxi,Aprile:2017iyp,Aprile:2015uzo} and indirect experiments\cite{Conrad:2014tla}.
 It is interesting to note 
that the presence of extra scalar in the form of DM can shift the instability scale ($\Lambda_I$) 
toward larger value compared to the SM  one
 ($\Lambda_I>\Lambda_I^{\textrm{SM}}$)\cite{Haba:2013lga,Khan:2014kba,Khoze:2014xha,Gonderinger:2009jp,Gonderinger:2012rd,Chao:2012mx
,Gabrielli:2013hma}.

On the other hand to accommodate non-zero neutrino mass via type-I seesaw mechanism,  one can extend SM with three
right handed(RH) neutrinos. The RH neutrinos, being SM singlet, will have standard Yukawa like coupling involving Higgs and lepton
doublets. The presence of the neutrino Yukawa coupling affects the running of the Higgs quartic coupling
similar to the top Yukawa coupling. In fact with neutrino Yukawa coupling, $Y_{\nu}$, of $\mathcal{O}(1)$, $\Lambda_I$ could 
be  lower than $\Lambda_I^{\textrm{SM}}$\cite{Chen:2012faa,Rodejohann:2012px,Rose:2015fua,Bambhaniya:2016rbb,Lindner:2015qva,
Chakrabortty:2012np,Datta:2013mta,Coriano:2014mpa,Ng:2015eia,Bonilla:2015kna,Khan:2012zw}, that might lead to an unstable 
Universe. The situation does not alter much even if one includes scalar singlet DM ($m_{\textrm{DM}}\leq 500$ GeV) in this framework \cite{Chao:2012mx,Chen:2012faa,
Garg:2017iva,Ma:2006km,Davoudiasl:2014pya,Chakrabarty:2015yia}. So the combined 
framework of RH neutrinos and scalar singlet DM excludes a significant range of DM mass ($< 500$ GeV) while keeping 
the EW vacuum on the verge of being unstable.

With an endeavour to make the EW vacuum absolutely stable upto the planck scale
$M_P$, in a scenario that can accommodate both the DM and massive 
neutrinos with large $Y_{\nu}$ (in type-I seesaw) and simultaneously 
to reopen the window for lighter 
scalar singlet DM mass ($<$ 500 GeV), we incorporate two SM real 
singlet scalars and three SM singlet RH neutrinos in this work.

Similar models to address DM phenomenology
involving additional scalars  (without involving
RH neutrinos) have been studied \cite{Abada:2013pca,Gabrielli:2013hma,Ahriche:2013vqa,Chakrabarty:2014aya,Das:2015mwa,Chakrabarty:2016smc}, however with different
motivations. Our set up also differs from them in terms of
 inclusion of light neutrino mass through type-I seesaw. The proposed model 
has several important ingredients which are mentioned below along with 
their importance. 

\begin{itemize}
\item One of the additional SM singlet scalars is our DM candidate whose
stability is achieved with an unbroken $Z_2$ symmetry.
 
\item The other scalar would acquire a nonzero vacuum expectation 
value (vev). This field has two fold contributions in our analysis: 
(i) it affects the running of the SM Higgs quartic coupling and (ii) the dark 
matter phenomenology becomes more involved due to its mixing with the SM Higgs
and the DM. 

\item The set up also contains three RH neutrinos in order to generate 
non-zero light neutrino mass through type-I seesaw mechanism.  Therefore, 
along with the contributions from the additional scalar fields, neutrino 
Yukawa coupling, $Y_{\nu}$, is also involved in studying the running of the 
Higgs quartic coupling.
\end{itemize}

We observe that the presence of the scalar\footnote{This scalar perhaps 
can be identified with moduli/inflaton fields \cite{Ema:2016ehh, Bhattacharya:2014gva,
Saha:2016ozn,Okada:2015zfa,EliasMiro:2012ay} or messenger field\cite{Baek:2012uj} 
connecting SM and any hidden sector.} with non-zero vev  affects the DM phenomenology in such a way that 
$m_{\textrm{DM}}$  less than 500 GeV  becomes perfectly allowed mass range 
considering the recent XENON-1T result\cite{Aprile:2017iyp}, which otherwise was excluded 
from the DM direct search results\cite{Arcadi:2017kky}. We also include XENON-nT\cite{Aprile:2017iyp} 
prediction to further constrain our model. On the other hand, we find that the SM Higgs quartic 
coupling may remain positive till $M_P$ (or upto some other scale higher than $\Lambda_I^{\textrm{SM}}$) 
even in presence of large $Y_\nu$, thanks to the involvement of the 
scalar with non-zero vev.  We therefore identify the relevant parameter 
space (in terms of stability, metastability and instability regions) of the 
model which can allow large $Y_{\nu}$ (with  different mass scales of RH neutrinos) and scalar 
DM below 500 GeV. Bounds from other related aspects, $e.g.$  lepton flavor 
violating decays, neutrinoless double beta decay etc., are also considered.  
The set-up therefore demands rich phenomenology what we present in the following 
sections.

The paper is organized as follows. In section 
\ref{EXHRH}, we discuss the set-up of our model and in section \ref{Constrain}, we include the constraints 
on our model parameters. Then in the subsequent sections \ref{DMPheno} and 
\ref{VSDMRH}, we discuss the DM phenomenology and vacuum stability respectively in the context of  our model. 
In section \ref{Nu}, we discuss connection of the model with other observables. Finally we 
conclude in section \ref{Conclu}.

\section{The Model}\label{EXHRH}

As mentioned in the introduction, we aim to study how the EW vacuum can be made stable in a model that 
would successfully accommodate a scalar DM and neutrino mass. For this purpose, we extend the SM by 
introducing two SM singlet scalar fields, $\phi$ and $\chi$, and three right-handed neutrinos, $N_{i=1,2,3}$. We 
have also imposed a discrete symmetry, $Z_2 \times Z'_2$.  The field $\phi$ is odd (even) under $Z_2$ ($Z'_2$) 
and $\chi$ is even (odd) under $Z_2$ ($Z'_2$) while all other fields are even under both $Z_2$ and $Z'_2$. 
There exists a non-zero vev associated with the $\chi$ field.  The unbroken $Z_2$ ensures the stability of our 
dark matter candidate $\phi$. On the other hand, the inclusion of $Z'_2$  simplifies the scalar  
potential in the set-up\footnote{A spontaneous breaking of discrete symmetry may lead to cosmological 
domain wall problem \cite{Dvali:1994wv}. To circumvent it, one may introduce explicit $Z'_2$ breaking term in higher 
order which does not affect our analysis.}.  The RH neutrinos are included in order to incorporate the light 
neutrino mass through type-I seesaw mechanism.

The scalar potential involving  $\phi, \chi$ and the SM Higgs doublet ($H$) is given by
\begin{align}
 V = V_\textrm{I} + V_{\textrm{II}}+ V_{\textrm{III}} + V_{\textrm{H}},
\label{eq:lag_tot}
\end{align}
where 
\begin{eqnarray}
V_{\rm{I}} &=& \frac{1}{2}  \mu_{\phi}^2 \phi^2
+ \frac{1}{4!} \lambda_{\phi} \phi^4 + \frac{1}{2} \lambda_{\phi H} \phi^2  H^{\dagger} H ; \nonumber\\
V_{\rm{II}} &=&-\frac{1}{2}\mu_\chi^2\chi^2+\frac{\lambda_\chi}{4!}\chi^4 + \frac{\lambda_{\chi H}}{2}\chi^2|H|^2; 
\nonumber\\
V_{\rm{III}} &=& \frac{1}{4} \lambda_{\chi \phi} \phi^2  \chi^2, ~~{\rm{and}} ~~
 V_{\rm{H}} =  -\mu^2_H H^{\dagger}H + \lambda_H (H^{\dagger}H)^2. \nonumber 
\end{eqnarray}
The relevant part of the Lagrangian responsible for neutrino mass is given by  
   \begin{equation}
   -\mathcal{L}_{\nu}=Y_{\nu_{ij}}\bar{l_L}_i\tilde{H}{N}_j+\frac{1}{2}{M_N}_{ij}N_i N_j, \nonumber 
  \end{equation}
where ${l_L}_i$ are the left-handed lepton doublets, $M_N$ is the Majorona mass matrix of the RH neutrinos. 
This leads to the light neutrino mass, $m_\nu=Y_{\nu}^T {M_N}^{-1} Y_{\nu} \frac{v^2}{2}$ with 
$v=246$ GeV as the vacuum expectation value of the SM Higgs.
 Minimization of the potential $V$ leads to the following vevs of $\chi$  and $H^0$ (the 
neutral component of $H$), as given by\footnote{Note that due to the absence of any $Z^\prime_2$ breaking term in the Lagrangian of
our model, panic vacua \cite{Barroso:2013awa,Barroso:2013kqa,Chakrabarty:2016smc} do not appear here. }
\begin{align}
v_\chi^2=6\frac{2\mu_\chi^2\lambda_H-\mu_H^2\lambda_{\chi H}}{2\lambda_H\lambda_\chi-3\lambda_{\chi H}^2}, \\
v^2=2\frac{\mu_H^2\lambda_\chi-3\mu_\chi^2\lambda_{\chi H}}{2\lambda_H\lambda_\chi-3\lambda_{\chi H}^2}.
\label{Pot1}
\end{align}

So after $\chi$ gets the vev and electroweak symmetry is broken, the mixing between 
$H^0$ and $\chi$ will take place and new mass or physical eigenstates,
 $H_1$ and $H_2$,
will be formed. The two physical eigenstates
are related with $H^0$ and $\chi$ by 
\begin{align}\label{eq:eigen}
 H_1=H^0 \cos\theta-\chi \sin\theta,\nonumber\\
 H_2=H^0 \sin\theta+\chi \cos\theta,
\end{align}
where the mixing angle $\theta$ is defined by
\begin{align}\label{tanth}
\tan2\theta=\frac{\lambda_{\chi H} v v_\chi}{-\lambda_H  v^2+\frac{\lambda_{\chi } v_{\chi}^2}{6}}. 
\end{align}
Similarly the mass eigenvalues of these physical Higgses are found to be
 \begin{align}
 m_{H_1}^2=\frac{\lambda_\chi}{6} v_\chi^2(1-\sec2\theta)+\lambda_Hv^2(1+\sec2\theta)\label{massE1},\\
 m_{H_2}^2=\frac{\lambda_\chi}{6} v_\chi^2(1+\sec2\theta)+\lambda_Hv^2(1-\sec2\theta)\label{massE2}.
 \end{align}
Using Eqs.(\ref{tanth},\ref{massE1},\ref{massE2}), the couplings $\lambda_H$, $\lambda_\chi$ 
and $\lambda_{\chi H}$ can be expressed in terms 
of the masses of the physical eigenstates $H_1$ and $H_2$, the vevs ($v$, $v_\chi$) and the mixing angle 
$\theta$ as 
\begin{align}
\lambda_H =&\frac{m_{H_1}^2}{4 v^2}(1+\cos 2\theta)+\frac{m_{H_2}^2}{4 v^2}(1-\cos2\theta)\label{lambdaH},\\
\lambda_\chi=&\frac{3m_{H_1}^2}{2 v_\chi^{2}}(1-\cos 2\theta)+\frac{3m_{H_2}^2}{2 v_{\chi}^ 2}(1+\cos2\theta)\label{lambdaChi},\\
\lambda_{\chi H}=&\sin2\theta\Big(\frac{m_{H_2}^2-m_{H_1}^2}{2 v v_\chi}\Big)\label{lambdaChiH}.
\end{align}
Among $H_1$ and $H_2$, one of them would be the Higgs discovered at LHC. The other 
Higgs can be heavier or lighter than the SM Higgs. Below we proceed to discuss 
the constraints to be imposed on the couplings and mass parameters of the model before
studying the DM phenomenology and vacuum stability in the subsequent sections.
\section{Constraints}\label{Constrain} 

Here we put together the constraints (both theoretical and 
experimental) that we will take into account to find the parameter space of our model. 

\begin{itemize}

\item  In order to keep the entire potential stable, one needs to maintain the following 
conditions involving the couplings present in $V$ (considering all couplings as real)
\begin{align}\label{stabC}
 \textrm{\textbf{ST$_{1,2,3}$: }}& \lambda_H>0,\textrm{   }\lambda_{\chi}>0, \textrm{   }\lambda_{\phi} > 0,\nonumber \\
\textrm{\textbf{ST$_{4,5,6}$: }}&\lambda_{\chi H}+\sqrt{\frac{2}{3}\lambda_H\lambda_\chi} > 0,\textrm{   }
 \lambda_{\phi H}+\sqrt{\frac{2}{3}\lambda_H\lambda_\phi}>0, \textrm{   }
3\lambda_{\chi\phi}+\sqrt{\lambda_\chi\lambda_\phi}>0, \nonumber\\
\textrm{\textbf{ST$_{7}$: }}&\sqrt{\lambda_H\lambda_\chi\lambda_\phi}+\lambda_{\chi H}\sqrt{\frac{3}{2}\lambda_\chi}
+3\lambda_{\phi H}\sqrt{\lambda_H}+3\lambda_{\chi\phi}\sqrt{\lambda_H},\nonumber\\
&+3\Big[\Big(\lambda_{\chi H}+\sqrt{\frac{2}{3}\lambda_H\lambda_\chi}\Big)
\Big(\lambda_{\phi H}+\sqrt{\frac{2}{3}\lambda_H\lambda_\phi}\Big)
\Big(\lambda_{\chi \phi}+\frac{1}{3}\sqrt{\lambda_\phi\lambda_\chi}\Big)\Big]^{1/2}>0,
\end{align}
which followed from the co-positivity of the mass-squared matrix involving $H, ~\chi$ 
and $\phi$\cite{Kannike:2012pe,Chakrabortty:2013mha}. 

\item In addition, the perturbative unitarity associated with the $S$ matrix corresponding to 
2 $\rightarrow$ 2 scattering processes involving all two particles initial and final states
 \cite{Horejsi:2005da,Bhattacharyya:2015nca} 
are considered. In the specific model under study, there are eleven neutral and four singly 
charged combinations of two-particle initial/final states. The details are provided in 
Appendix A. It turns out that the some of the scalar couplings of Eq.(\ref{eq:lag_tot}) 
are bounded by 
\begin{eqnarray}\label{puc}
 \lambda_{H} ~ < 4 \pi ,~~~~  \lambda_{\phi H} ~ < 8 \pi ,~~~~  \lambda_{\chi H} ~ < 8 \pi ,
 ~~~~  \lambda_{\chi \phi} ~ < 8 \pi\rm.
\end{eqnarray}  
The other scalar couplings are restricted (in form of combinations among them) from the 
condition that the roots of a polynomial equation should be less than 16$\pi$ (see Eq.(\ref{roots}) 
of Appendix A). 

\item To maintain the perturbativity of all the couplings, we impose the condition that the scalar 
couplings should remain below 4$\pi$  while Yukawa couplings are less than $\sqrt{4\pi}$  till 
$M_{P}$. An upper bound on $\tan\beta(=v/v_\chi)$ follows from the perturbativity of $\lambda_{\chi}$\cite{Lopez-Val:2014jva} 
with a specific choice of $m_{H_2}$.

\item Turning into the constraints obtained from experiments, we note that the observed signal strength of 
the 125 GeV Higgs boson at LHC \cite{Khachatryan:2015cwa,Aad:2015kna,Chatrchyan:2013mxa,CMS1,CMS2,Strassler:2006ri}
 provides a limit on $\sin\theta$ as, 
$|\sin\theta| \lesssim 0.36$ with $m_{H_2} \gtrsim 150$ GeV.  The analysis in \cite{Robens:2016xkb} 
shows that  $\sin\theta$ is restricted significantly ($|\sin\theta| \lesssim 0.3$) by the direct Higgs 
searches at colliders \cite{Khachatryan:2015cwa,Aad:2015kna,Chatrchyan:2013mxa,CMS1,CMS2} and combined Higgs 
signal strength \cite{ATLAS1} for 150 GeV $<m_{H_2} < 300$ GeV while for 300 GeV $<m_{H_2} <$ 800 GeV, 
it is the NLO contribution to the $W$ boson mass \cite{Lopez-Val:2014jva} which restricts $\sin\theta$ 
in a more stipulated range. Corrections to the electroweak precision observables through the $S, T, U$ parameters turn 
out to be less dominant compared to the limits obtained from $W$ boson mass correction \cite{Lopez-Val:2014jva}.
For our purpose, we consider $\sin\theta \lesssim 0.3$ for the analysis.

\end{itemize}  

Apart from these, we impose the constraints on $Y_\nu$ from lepton flavor violating decays. Also 
phenomenological limits obtained on the scalar couplings involved in order to satisfy the relic density $(0.1175\leq\Omega h^2\leq 0.1219)$\cite{Ade:2013zuv} 
and direct search limits\cite{Aprile:2017iyp} by our dark matter candidate $\phi$ are considered when stability of 
the EW minimum is investigated. 

\section{Dark matter phenomenology}\label{DMPheno}
The scalar field $\phi$ playing the role of dark matter has a mass  
given by $m^2_{\textrm{DM}}=(\mu_{\phi}^2+
\frac{1}{2}\lambda_{\phi H}v^2)$  as followed from Eq.(\ref{eq:lag_tot}). Before moving toward the relic density 
calculation in our model, we would like to comment on the simplest $Z_2$ odd scalar dark matter 
scenario in view of recent XENON 1T\cite{Aprile:2017iyp} result. Note that for the purpose of DM phenomenology in 
this case, the only relevant parameters are given by $m_{\textrm{DM}}$ and the Higgs portal coupling 
$\lambda_{\phi H}$ (or $\mu_\phi$ and $\lambda_{\phi H}$).  
\begin{figure}[h]
$$
\includegraphics[height=5.0cm]{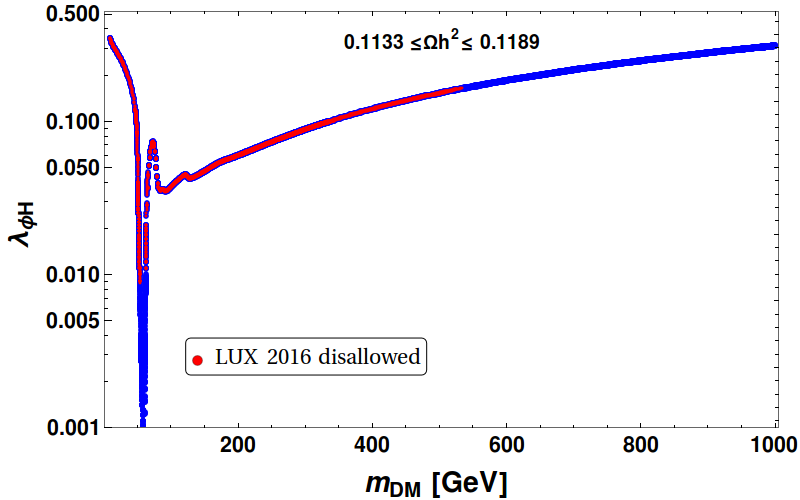}
\includegraphics[height=5.0cm]{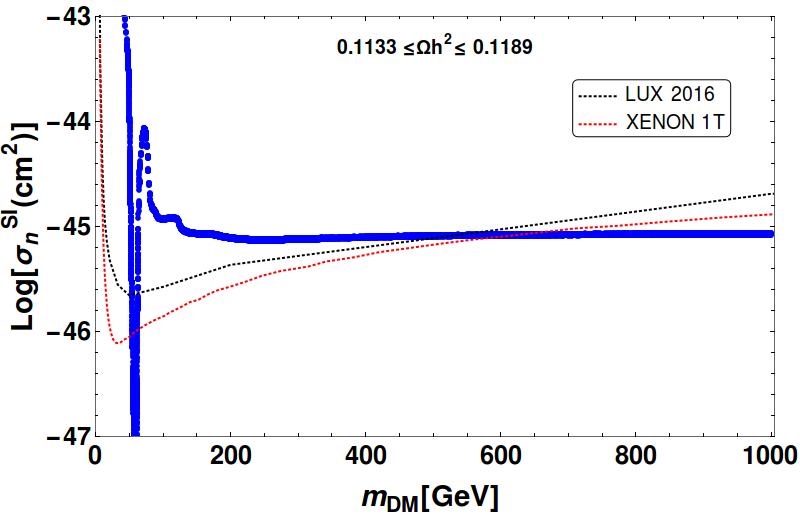}
 $$
 \caption{(Left panel:) 
 relic density contour plot in $m_{\textrm{DM}}-\lambda_{\phi H}$ plane; the red portion corresponds to the disfavored range
 of parameters by recent direct detection results, while the blue portion stands for the allowed region of parameters consistent 
with direct detection results. (Right panel:) Spin independent cross section is plotted (blue line) for relic density allowed points as a function 
 of $m_{\textrm{DM}}$, where the LUX 2016 and XENON 1T limits are indicated by blue and red dashed lines respectively.}
 \label{fig:sigmav-compare1}
 \end{figure}
In Fig.\ref{fig:sigmav-compare1} 
(left panel), we provide a contour plot for relic density consistent with the Planck result\cite{Ade:2013zuv} in the
 $\lambda_{\phi H}-m_{\textrm{DM}}$ plane indicated by the blue solid line. In the right panel of Fig.\ref{fig:sigmav-compare1},
we provide the DM-nucleon cross section evaluated with 
the value of $\lambda_{\phi H}$ corresponding to the $m_{\textrm{DM}}$ value as obtained from the left 
panel plot.
 We then incorporate the direct search limits on the DM-nucleon cross 
section as obtained from LUX 2016 \cite{Akerib:2016vxi}, and the most recent XENON 1T\cite{Aprile:2017iyp} result 
\cite{Aprile:2017iyp} in the same plot denoted by blue and red dashed lines respectively.
 We conclude that the dark matter mass below 500 GeV is excluded from the present 
XENON 1T\cite{Aprile:2017iyp} result. This result is indicated by the red portion of the contour line in the
left panel, while the remaining blue portion of the contour plot ( of the left panel)
represents the allowed range of $m_{\textrm{DM}}$ satisfying both the relic density and direct search 
constraints. 

\begin{figure}
 \begin{center}
   \includegraphics[width=14cm, height=7cm]{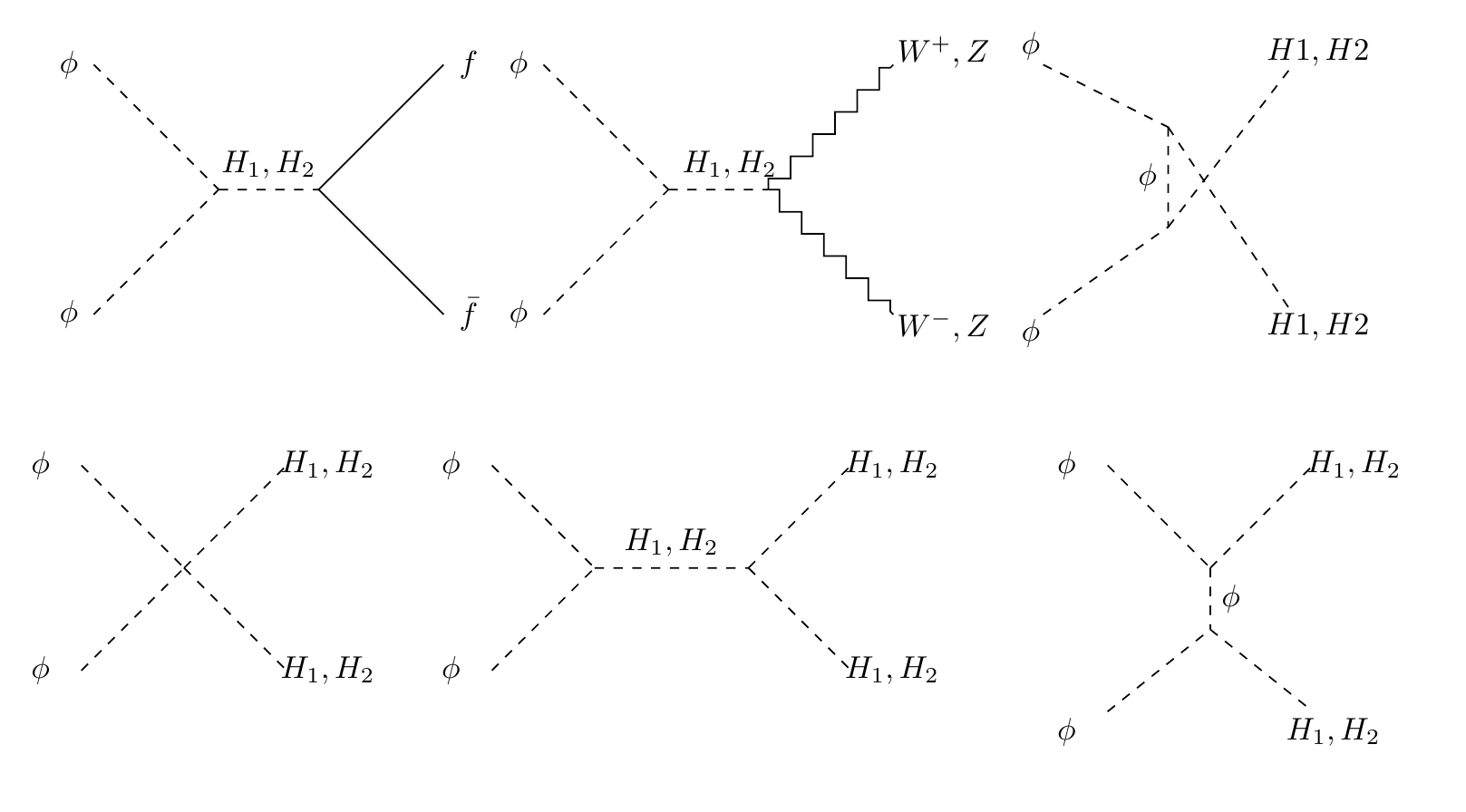}
 \end{center}
\caption{ Diagrams contributing to $\phi \phi$ annihilation to SM particles and the other Higgs. }
\label{ann_diag}
 \end{figure}

Let us now move to the relic density estimate in our set-up with the extra scalar $\chi$ and compare the 
phenomenology with the simplest scalar DM scenario in the light of the mixing between the SM Higgs and $\chi$. 
Using Eq.(\ref{eq:eigen}) and inserting them into the SM Lagrangian along with the ones mentioned in 
Eq.(\ref{eq:lag_tot}), we obtain the following list of interaction vertices involving two 
Higgses ($H_1$ and $H_2$), dark matter field ($\phi$) and several other SM fields.

\begin{eqnarray}\label{Coups}
  H_1 f \bar{f}, H_2 f \bar{f}  & :&  \frac{m_f}{v} \cos\theta , \frac{m_f}{v} \sin\theta \nonumber \\
  H_1 Z Z, H_2 Z Z & :&  \frac{2 m_Z^2}{v} \cos\theta g^{\mu \nu}, \frac{2 m_Z^2}{v} \sin\theta g^{\mu \nu} \nonumber \\  
  H_1 W^+W^-, H_2 W^+ W^- & :&  \frac{2 m_W^2}{v} \cos\theta g^{\mu \nu}, \frac{2 m_W^2}{v} \sin\theta g^{\mu \nu} \nonumber \\  
 \phi \phi H_1   & :&   -v_{\chi}\lambda_{\chi \phi} \sin\theta+ v \lambda_{\phi H} \cos\theta \equiv \lambda_{1}\nonumber \\
 \phi \phi H_2   & :&    v_{\chi}\lambda_{\chi \phi} \cos\theta+ v \lambda_{\phi H} \sin\theta \equiv \lambda_{2} \nonumber \\
 \phi \phi H_1 H_1 &:& \lambda_{\phi H} \cos^2\theta + \lambda_{\chi \phi} \sin^2\theta\nonumber \\
 \phi \phi H_2 H_2 &:& \lambda_{\phi H} \sin^2\theta + \lambda_{\chi \phi} \cos^2\theta\nonumber \\
 \phi \phi H_1 H_2 &:& (\lambda_{\phi H} - \lambda_{\chi \phi} )\sin\theta \cos\theta\nonumber \\
 H_1 H_1 H_1  &:& [6 v \lambda_H \cos^3\theta -3 v_\chi \lambda_{\chi H} \cos^2\theta \sin\theta +3 v \lambda_{\chi H} \cos\theta \sin^2\theta-v_\chi \lambda_\chi \sin^3\theta]\nonumber \\
 H_2 H_2 H_2  &:& [6 v \lambda_H \sin^3\theta +3 v_\chi \lambda_{\chi H} \cos\theta \sin^2\theta +3 v \lambda_{\chi H} \cos^2\theta \sin\theta+v_\chi \lambda_\chi \cos^3\theta]\nonumber \\
 H_1 H_1 H_2  &:& [2 v (3 \lambda_H - \lambda_{\chi H})\cos^2\theta \sin\theta +v \lambda_{\chi H} \sin^3\theta + v_\chi\nonumber (\lambda_{\chi}-2\lambda_{\chi H})\cos\theta \sin^2\theta 
 \\ &&+v_\chi \lambda_{\chi H}\cos^3\theta ]\nonumber \\
 H_1 H_2 H_2  &:& [2 v (3 \lambda_H - \lambda_{\chi H})\cos\theta \sin^2\theta +v \lambda_{\chi H} \cos^3\theta - v_\chi (\lambda_{\chi}-2\lambda_{\chi H})\cos^2\theta \sin\theta\nonumber
 \\&& -v_\chi \lambda_{\chi H}\sin^3\theta ].\nonumber \\
\end{eqnarray}
Following Eq.(\ref{Coups}) we draw the Feynman diagrams for DM annihilation channels into SM particles and
 to the second Higgs in Fig.\ref{ann_diag}.

It is expected that the DM candidate is in thermal equilibrium with the SM degrees of freedom in the 
early universe. We therefore proceed to evaluate their abundance through the standard freeze-out 
mechanism. The Boltzmann equation, 
\begin{eqnarray}
&& \dot{n}_{\phi}+3\textrm{H} n_{\phi} =-\langle \sigma v _{ \phi \phi} \rangle
\left(n_{\phi}^2- {n_{\phi}^{eq}}^2\right),
\label{beq2}
\end{eqnarray}
is employed for this purpose, where $n_{\phi}$ is the number density of the dark matter $\phi$, H is the 
Hubble parameter, 
$\langle \sigma v_{\phi \phi} \rangle$ represents the total annihilation cross-section as given by  
$\langle \sigma v_{\phi \phi} \rangle = \langle \sigma v _{\phi \phi \to SM,SM}
 \rangle+\langle \sigma v _{ \phi \phi \to H_1 H_2} \rangle
+ \langle \sigma v _{ \phi \phi \to H_2 H_2} \rangle$.
We consider here the RH neutrinos to be massive enough
compared to the DM mass. So RH neutrinos do not participate in DM phenomenology.
We have then used the MicrOmega package\cite{Belanger:2013oya} to evaluate the final relic abundance of DM. 

We have the following parameters in our set-up,
\begin{align}\label{Allpara}
\{m_{H_1}, m_{H_2}, m_{\textrm{DM}}, \sin\theta, \lambda_{\chi \phi}, \lambda_{\phi H}, v,\tan\beta, \lambda_{\phi}\}. 
\end{align}
The parameters $v_{\chi}$ is involved in the definition of $\tan\beta = v/v_{\chi}$. Parameters 
$(\lambda_{H}, \lambda_{\chi}, \lambda_{\chi H})$ can be written in terms of other parameters as shown in
 Eqs.(\ref{lambdaH},\ref{lambdaChi},\ref{lambdaChiH}). 
Among all the parameters in Eq.(\ref{Allpara}), $\lambda_{\phi}$ does not play any significant role in DM analysis.

We first assume $H_1$ as the Higgs discovered at LHC {\it{i.e.}} $m_{H_1}\sim$125.09 GeV\cite{Agashe:2014kda} and the other 
Higgs is the heavier one ($m_{H_2}>m_{H_1}$).
It would be appealing in view of LHC accessibility to keep $m_{H_2}$ below 1 TeV. In this case limits on $\sin\theta, \tan\beta$ are applicable as discussed
in Section \ref{Constrain} depending on 
specific value of $m_{H_2}$ \cite{Robens:2016xkb}. Now in this regime (where $m_{H_2}$ is not too heavy, in particular $m_{H_2}<1$ TeV),
 $\sin\theta$ is bounded by $\sin\theta \lesssim 0.3$ \cite{Robens:2016xkb}
and we have taken here a conservative choice by fixing $\sin\theta = 0.2$.
Note 
that in the small $\sin\theta$ approximation, $H_1$
 is mostly dominated by the SM Higgs doublet $H$. In this limit the second term in Eq.(\ref{lambdaH}) effectively provides the threshold correction
to $\lambda_H$\cite{Basso:2013nza,EliasMiro:2012ay,Lebedev:2012zw} which helps in achieving vacuum stability as we 
will see later. Furthermore considering this threshold effect to be equal or less 
than the first term in Eq.(\ref{lambdaH}) ({\it{i.e.}} approximately the SM value of $\lambda_H$),
 we obtain an  upper bound on $m_{H_2}$ as $m_{H_2}<\frac{m_{H_1}}{\tan\theta}$.
 Therefore in case with $m_{H_2}>m_{H_1}$, our working regime of $m_{H_2}$ can be considered within 
  $\frac{m_{H_1}}{\tan\theta}>m_{H_2}>m_{H_1}$. We take $m_{H_2}$ to be 300 GeV for our analysis.
  
Note that with small $\theta$,
 $\lambda_{\chi}$ almost coincides with the second term in Eq.(\ref{lambdaChi}).
 It is quite natural to keep the magnitude 
 of a coupling below unity to maintain the perturbativity limit for all energy scales including its running.
Hence with the demand $\lambda_\chi<1$, one finds $v_\chi>\sqrt{3}m_{H_2}$. 
 To show it numerically, let us choose $\sin\theta=0.2$, then we obtain $125\textrm{ GeV}<m_{H_2}<620$ GeV. 
 Therefore with $m_{H_2}=300$ GeV, a lower limit on $v_\chi\geq520$ GeV can be set. We consider $v_{\chi}$ to 
 be 800 GeV so that $\tan\beta$ turns out to be 0.307. 
 
 On the other hand, if we consider the other Higgs
 to be lighter than the one discovered at LHC, we identify $m_{H_2}$ to be the one found at LHC and hence $m_{H_1}\le 125$ GeV. Then
  Eq.(\ref{eq:eigen}) suggests $\sin\theta\rightarrow 1$ as the complete decoupling limit of the second Higgs. 
  Following the analysis
  in \cite{Robens:2015gla,Robens:2016xkb,Chalons:2016jeu,Dawson:2015haa,Fischer:2016rsh,Lewis:2017dme}, we infer that 
  most of the parameter space except for a very narrow
  region both in terms of mixing angle ($\sin\theta\sim 0.9$) and mass of the lighter Higgs  ($m_{H_1}\sim 85-100$) GeV,
 is excluded from LEP and LHC
  searches. 
  \begin{figure}[h]
 \begin{center}
 \includegraphics[width=4.5cm, height=4.5cm]{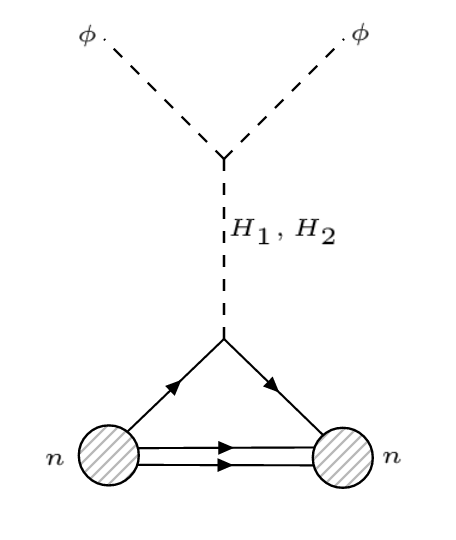}
 \end{center}
\caption{Feynman diagram for DM Direct Detection.}
  \label{fig:DD}
 \end{figure}
  Such a range is not suitable for our purpose as can bee seen from Eq.(\ref{lambdaH}). 
  In this large $\sin\theta$ limit,
 $\lambda_H$ gets the dominant contribution from the second term in Eq.(\ref{lambdaH}) where the first term serves the
  purpose of threshold effect on $\lambda_H$. However $m_{H_1}$ being smaller than $m_{H_2}$ (the SM like Higgs), 
  this effect would not be sufficient to enhance $\lambda_H$ such that its positivity till $M_P$ can be ensured.
 Therefore we discard the scenario $m_{H_1}<m_{H_2}$ (SM like Higgs)
    from our discussion. Hence the DM phenomenology basically depends on $m_{\textrm{DM}}, \sin\theta, \lambda_{\chi \phi}$ and 
 $\lambda_{\phi H}$.

In a direct detection experiment, the DM scatters with the nucleon through the exchange of $H_1$ 
and $H_2$ as shown schematically in Fig.\ref{fig:DD} .
 The resulting spin-independent cross-section of DM-nucleon elastic scattering is given by \cite{Gabrielli:2013hma} :
  \begin{eqnarray}\label{DD_formulla}
  \sigma_{n}^{SI}=\frac{f_n^2 \mu_{n}^2 m_n^2}{4\pi v^2 m_{\textrm{DM}}^2} \Big[\frac{\lambda_{1} \cos\theta}{m_{H_1}^2}
+\frac{\lambda_{2} \sin\theta}{m_{H_2}^2}\Big]^2,
  \end{eqnarray}
 where $\mu_n=\frac{m_n m_{\textrm{DM}}}{m_n+m_{\textrm{DM}}}, f_n=0.284$ \cite{Alarcon:2011zs,Alarcon:2012nr}. The couplings appeared as $\lambda_{1}, 
 \lambda_{2}$ are specified in the list of vertices in Eq.(\ref{Coups}). Below we discuss how we can  estimate the relevant parameters 
($\lambda_{\phi_H}$,$\lambda_{\chi\phi}$ and $m_{\textrm{DM}}$) from 
 relic density and direct search limits. For this purpose, we consider $m_{H_2}=300$ GeV and $v_\chi=800$ GeV as reference values, unless
  otherwise mentioned. 
\subsection{DM mass in region R1: [$150\textrm{ GeV}<m_{\textrm{DM}}\leq 500$ GeV]}
\begin{figure}[htb!]
 \begin{center}
 \includegraphics[width=7.3cm,height=6cm]{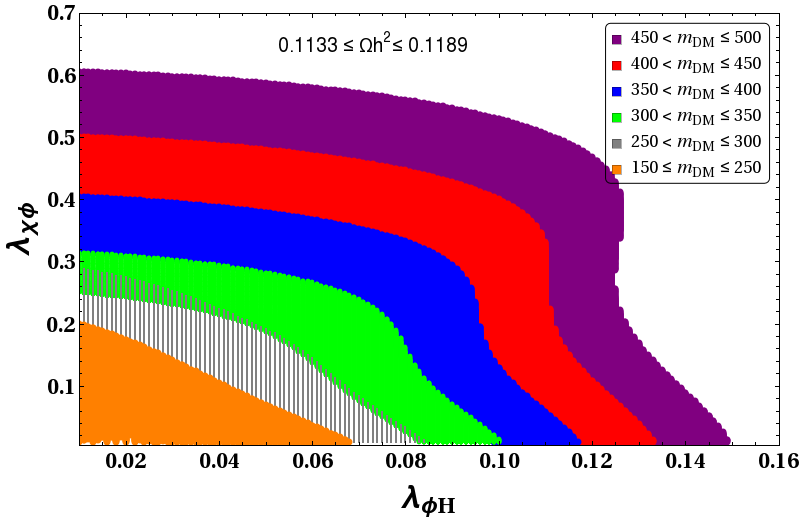}
  \includegraphics[width=7.3cm,height=6cm]{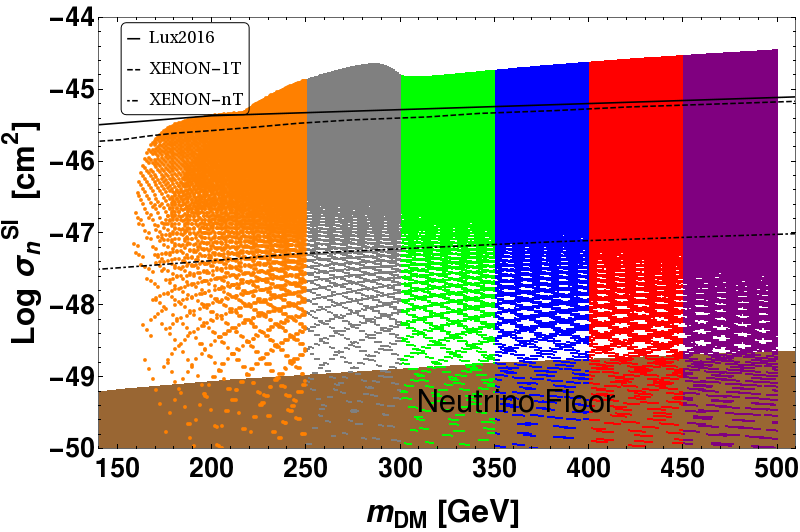}\\
 \includegraphics[width=7.3cm,height=6cm]{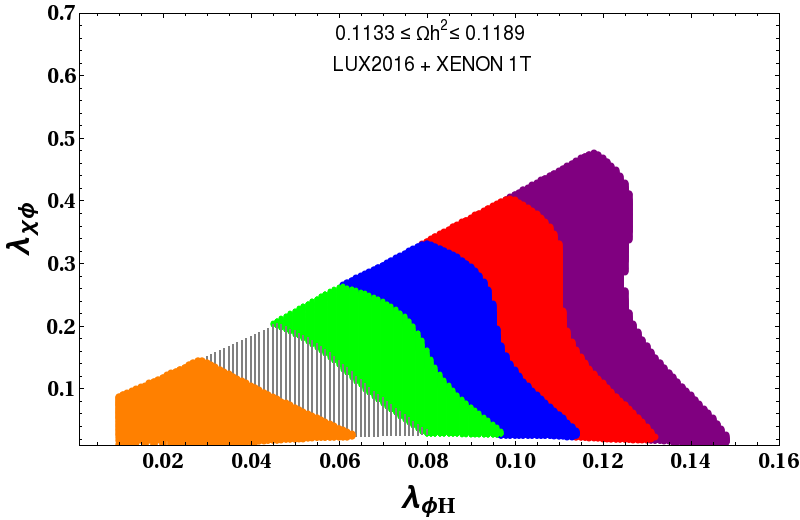}
  \includegraphics[width=7.3cm,height=6cm]{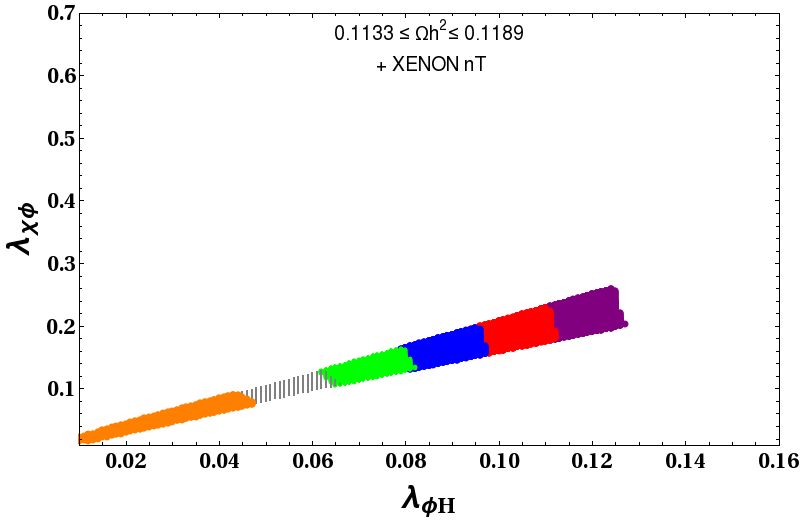}\\ 
 \end{center}
\caption{Top left: Allowed points on $\lambda_{\phi H}$-$\lambda_{\chi\phi}$ plane for DM having mass
 $150<m_{\textrm{DM}}<500$ GeV to satisfy correct order of relic density. Top right: Spin independent nucleon 
  cross section of DM has been plotted against the DM mass. Bottom panel: The top left plot has been 
  constrained using recent LUX 2016\cite{Akerib:2016vxi}, Xenon 1T\cite{Aprile:2017iyp} limits to
produce bottom-left figure and Xenon nT\cite{Aprile:2015uzo} predictions to get bottom-right figure.}
  \label{correlation}
 \end{figure}

In this region any decay mode
of $H_1$ and $H_2$ into DM is kinematically forbidden following our consideration for $m_{H_2}=300$ GeV. 
As stated before, we consider $m_{H_1}$ to be the SM like Higgs discovered at LHC, with
 $v_\chi=800$ GeV and $\tan\beta$ is fixed at 0.307. Therefore in order to
 satisfy the relic density $\Omega h^2=0.1161 \pm 0.0028$\cite{Ade:2013zuv}, we first scan over 
$\lambda_{\phi H}$ and $\lambda_{\chi \phi}$ for different ranges of dark matter mass 
where $\sin\theta$ is kept fixed at 0.2. The allowed range of parameter space
 contributing to the relic abundance satisfying the correct
relic density is indicated
on $\lambda_{\phi H}-\lambda_{\chi \phi}$ plane in Fig.\ref{correlation} (in the top left 
panel), where different coloured patches indicate different ranges of $m_{\textrm{DM}}$. 
 In the upper-right plot of Fig.\ref{correlation}, the corresponding direct search cross sections 
 for the relic density  
satisfied points obtained from the upper left plot (including the variation of $\lambda_{\phi H}, \lambda_{\chi \phi}$) are provided.
It can be clearly seen that many of these points lie below the
LUX 2016\cite{Akerib:2016vxi} experimental limit for 
a wide range of dark matter mass (indicated by the colors depicted in the inset of  
Fig.\ref{correlation}, upper left panel).

  From the top left panel of Fig.\ref{correlation}, the relic density contour plot (with 
  a particular $m_{\textrm{DM}}$) in $\lambda_{\chi \phi}$-$\lambda_{\phi H}$ plane shows that there exists a 
  range of $\lambda_{\phi H}$ for which the plot is (almost) insensitive to the change in $\lambda_{\chi 
  \phi}$. This becomes more prominent for plots associated with higher dark matter mass. In particular, 
  the contour line satisfying the correct  relic density with $m_{\textrm{DM}} = 500$ GeV depicts a sharp variation 
  in $\lambda_{\chi\phi}$ (below 0.4) with almost no variation of $\lambda_{\phi H}$ around 0.13. We 
  now discuss the reason behind such a behaviour.  We note that for $\lambda_{\phi H} > 0.13$, the 
  total annihilation cross section satisfying the relic density is mostly dominated by the $\phi\phi\rightarrow$ 
  SM, SM process, specifically $\phi\phi \rightarrow WW, ZZ$ dominate. In our scenario, 
  $\phi\phi \rightarrow WW, ZZ$  processes are mediated by both the Higgses, $H_1$ and $H_2$. Although 
  $\lambda_{\chi \phi}$ is involved in the vertices characterizing these processes, it turns out that once 
  both the $H_1, H_2$ contributions are taken into account, the $\lambda_{\chi \phi}$ dependence is effectively 
  canceled leaving the  $\phi\phi \rightarrow WW, ZZ$  annihilation almost independent of $\lambda_{\chi \phi}$. 
  Hence $\phi\phi \rightarrow$ SM, SM depends mostly on $\lambda_{\phi H}$. 
  The other processes like  $\phi\phi \rightarrow H_1H_2 (H_2H_2)$ are subdominant (these are allowed provided 
  $m_{\textrm{DM}} > 212.5 (300)$ GeV) in this region with large $\lambda_{\phi H}$. Then the total cross section 
  $\langle\sigma v_{\phi\phi}\rangle$ and hence 
  the relic density contour line becomes insensitive to the change in $\lambda_{\chi \phi}$ as long as it remains 
  below 0.4 while $\lambda_{\phi H} > 0.13$. This is evident in the top left panel of Fig.\ref{correlation}.  
  Similar effects are seen in case of lower $m_{\textrm{DM}}$ ($< 500$ GeV) as well.

    Once we keep on decreasing 
  $\lambda_{\phi H}$ below 0.13, it turns out that  $\phi\phi \rightarrow$ SM, SM becomes less important 
  compared to the  $\phi\phi \rightarrow H_2H_2$ (in particular the $t$ channel) with $\lambda_{\chi \phi}$ beyond 0.4 (in case of $m_{\textrm{DM}} 
  = 500$ GeV). Note that the plot shows the insensitiveness related to $\lambda_{\phi H}$ in this low $\lambda_{\phi H}$  region 
  for obvious reason. Similar results follow 
  with $m_{\textrm{DM}} < 300$ GeV also, where $\phi\phi \rightarrow H_1H_2$ provides the dominant contribution 
  in $\langle\sigma v_{\phi\phi}\rangle$. Based on our discussion so far we note that 
for $\lambda_{\chi\phi}\gg\lambda_{\phi H}$
 the channels with Higgses in the final states contribute more to total $\langle\sigma v_{\phi\phi}\rangle$.
  On the other hand for low values of $\lambda_{\chi\phi}$ (although comparable to $\lambda_{\phi H}$), the model resembles the usual Higgs portal
 dark matter scenario where W bosons in the final state dominate. To summarize,
\begin{itemize}
\item{ \textbf{$150 \textrm{ GeV}< m_{\textrm{DM}} < 212.5$~ GeV:} For low $\lambda_{\chi\phi}$, $\phi\phi \rightarrow W^+ W^-$ dominates.
 However  for large $\lambda_{\chi\phi}$, $\phi \phi \rightarrow H_1 H_1$ becomes the main annihilation channel.}
\item{\textbf{ $212.5 \textrm{ GeV}< m_{\textrm{DM}} < 300$~ GeV:} New annihilation process $\phi \phi \rightarrow H_1 H_2$ opens up.
 This with $\phi\phi\rightarrow H_1H_1$ contribute dominantly for large $\lambda_{\chi\phi}$. Otherwise the channels with SM
 particles in
 final states dominate.}
 \item{\textbf{$300 \textrm{ GeV}< m_{\textrm{DM}} < 500$~ GeV:} The annihilation channel $\phi \phi \rightarrow H_2
  H_2$ opens up in addition to $H_1H_1$ and $H_1H_2$ in the final states. Their relative contributions to
 total $\langle\sigma v_{\phi\phi}\rangle$ again depend on the value of
  $\lambda_{\chi\phi}$. }
 \end{itemize}

\begin{figure}[htb!]
 \begin{center}
 \includegraphics[width=7.3cm,height=6cm]{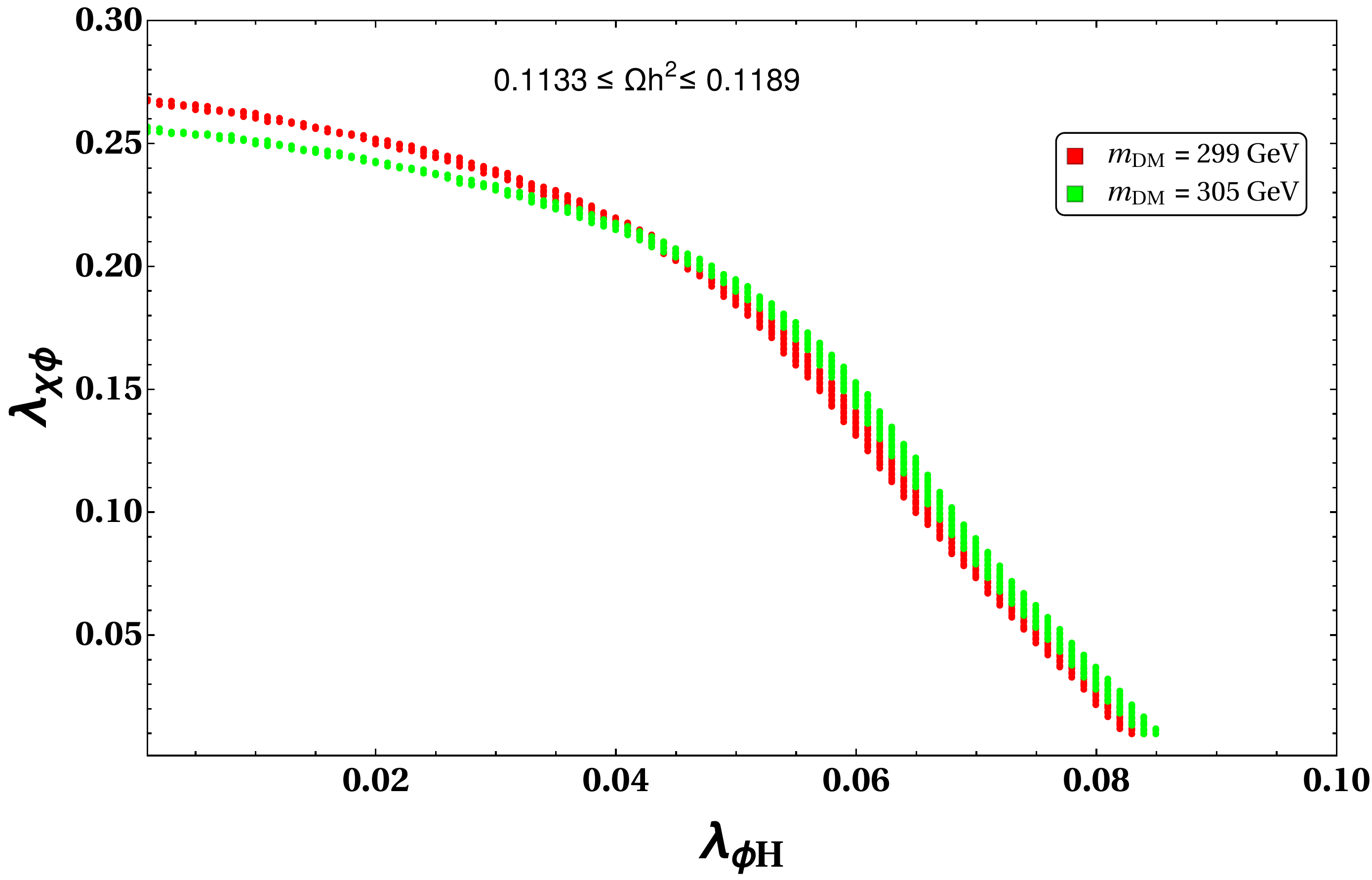}
 \end{center}
\caption{DM relic density contour lines in $\lambda_{\phi H}$-$\lambda_{\chi\phi}$ plane with $m_{\textrm{DM}}=299$ (red), 
305 GeV (green).}
  \label{Overlap}
 \end{figure}
In the top left panel of Fig.\ref{correlation}, we also note the existence of 
a small overlapped region when $\lambda_{\phi H}\ll\lambda_{\chi\phi}$ for the dark matter 
mass regions between 280-300 GeV and 300-310 GeV. This has been further clarified in Fig.\ref{Overlap},
 where we note that relic density contour lines with $m_{\textrm{DM}}=299$ 
GeV and $m_{\textrm{DM}}=305$ GeV intersect each other around $\lambda_{\phi H}\sim 0.05$ and 
$\lambda_{\chi\phi}\sim 0.21$. 
Note that when DM mass $m_{\textrm{DM}} \geq m_{H_2}=300$ GeV, in addition 
 to the $\phi\phi\rightarrow SM, SM$ and $\phi\phi\rightarrow H_1H_2$
 annihilation processes, $\phi\phi\rightarrow H_2 H_2$ opens up and contribute to the
  total annihilation cross section  
 ( this new channel can be realized through both $H_1$
  and $H_2$ mediation). 
  
Then total annihilation cross section will be enhanced for $m_{\textrm{DM}} > 300$ GeV case,
 {\it{i.e}} $\langle\sigma v_{\phi\phi}\rangle=\langle\sigma v\rangle_{\phi\phi\rightarrow SM, SM}+
 \langle\sigma v\rangle_{\phi\phi\rightarrow H_1,H_2}+\langle\sigma v\rangle_{\phi\phi\rightarrow H_2, H_2}$ becomes 
 large compared to the $280\textrm{ GeV}<m_{\textrm{DM}}<300$ GeV mass range where $\langle\sigma v\rangle_{\phi\phi\rightarrow H_2 H_2}$ 
 is not present. This enhancement has to be nullified in order  to realize the correct relic density and this is achieved 
 by reducing $\lambda_{\chi\phi}$ compared to its required value for a fixed $\lambda_{\phi H}$ and $m_{\textrm{DM}}$ in 
 $280\textrm{ GeV}\leq m_{\textrm{DM}}< 300$ region. Note that in view of our previous discussion, we already understand 
 that $\phi\phi\rightarrow H_2 H_2$ becomes important compared to $\phi\phi\rightarrow$ SM, SM process 
 in the region with $\lambda_{\chi \phi} \gg \lambda_{\phi H}$. Hence the two mass regions (below and above 300 GeV) 
 overlap  in $\lambda_{\phi H}-\lambda_{\chi\phi}$ plane as seen in the top left panel of  Fig.\ref{correlation} as well in Fig.\ref{Overlap}. 
 The total annihilation cross section of DM depends on its mass also.
 However the small mass differences between the two overlapped regions have very mild effect on $\langle\sigma v\rangle
   _{\textrm{Tot}}$.   
   Similar effect should be observed below and above $m_{\textrm{DM}} \sim (m_{H_1}+m_{H_2})/2 = $ 
  212.5 GeV as $\phi\phi\rightarrow H_1 H_2$ opens up there. However we find that around the $m_{\textrm{DM}}= 212.5$ GeV,
   even with $\lambda_{\chi\phi}\gg\lambda_{\phi H}$, the contribution from this particular channel to $\langle\sigma v\rangle
   _{\textrm{Tot}}$ is negligible as compared to  $\phi\phi\rightarrow$ SM SM contribution and hence we do not observe 
   any such overlapped region there.

  In the top right panel of Fig.\ref{correlation} we provide the spin-independent (SI) direct detection (DD) cross sections corresponding 
  to the points in the left panel satisfying relic density data having different range of dark matter 
   masses as indicated by the colored patches. We further put the LUX 2016\cite{Akerib:2016vxi}, XENON 1T\cite{Aprile:2017iyp} 
   and nT (expected) lines on it.
    As known, for a lowerer cross section, it reaches the neutrino floor where signals from DM can not be
  distinguished from  that of neutrino.      
We find that the scenario works with reasonable values of the parameters, {\it{i.e.}} not with 
any un-naturally small or large values of couplings. 
Note that once we use the XENON 1T\cite{Aprile:2017iyp} and projected XENON nT\cite{Aprile:2015uzo} limits on the scattering 
cross section, we would obtain more restricted region of parameter space for  
$\lambda_{\phi H}-\lambda_{\chi \phi}$ as shown in left (with XENON 1T\cite{Aprile:2017iyp}) 
and right (with XENON nT\cite{Aprile:2015uzo}) figures of the bottom panel. From the plot with XENON-nT prediction, 
we find that the scenario works even with reasonably large values of  $\lambda_{\phi H}, ~
\lambda_{\chi \phi}$ required for satisfying the relic density, although they are comparable 
to each other. This is because of the fact that to keep the direct detection cross section relatively 
small (even smaller than the XENON nT), it requires
 a cancellation between $\lambda_{\phi H}$ and $\lambda_{\chi\phi}$  
 as can be seen from Eq.(\ref{DD_formulla}) in conjugation with definition of $\lambda_1$ and $\lambda_2$ for a
  specific $\sin\theta = 0.2$ value. Such a cancellation is not that important 
for plots with LUX 2016\cite{Akerib:2016vxi} or XENON 1T\cite{Aprile:2017iyp} results and hence showing a wider region of parameter space for $\lambda_{\chi\phi}$
 and $\lambda_{\phi H}$. 

\begin{figure}[htb!]
 \begin{center}
 \includegraphics[width=8.5cm,height=5.8cm]{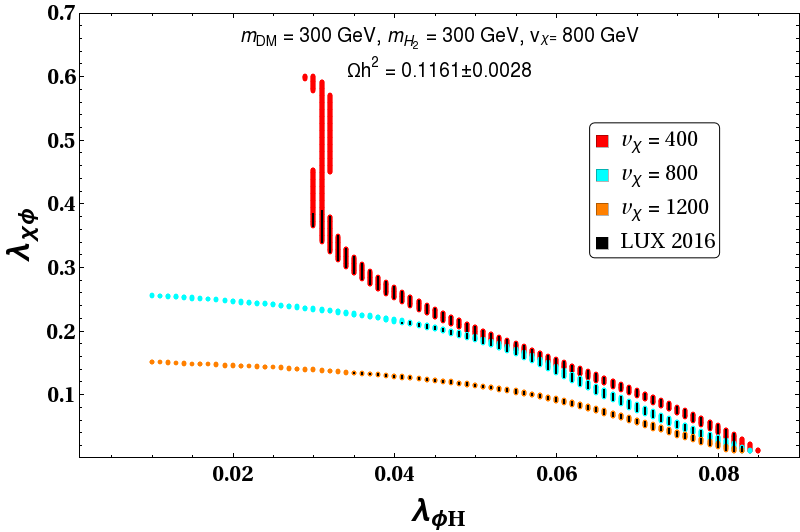}
 \end{center}
\caption{Allowed parameter space  to satisfy correct relic abundance  
 in $\lambda_{\phi H}-\lambda_{\chi\phi}$ plane with different $v_\chi$
 for $m_{\textrm{DM}}=300$ GeV . Other parameters $m_{H_2}=300$ GeV and 
 and $\sin\theta=0.2$ have been kept fixed. The LUX 2016\cite{Akerib:2016vxi} allowed region are also accommodated (solid black region) in the figures. }
  \label{fig:vchi}
 \end{figure}

 It can be concluded from upper panel of 
 Fig.\ref{correlation} that the presence of additional singlet scalar field $\chi$ 
 helps in reducing the magnitude of $\lambda_{\phi H}$ 
 that was required (say $\lambda^0_{\phi H}$)  to produce correct  relic density in minimal form
 of singlet 
 scalar DM or in other words it dilutes the pressure on $\lambda_{\phi H}$ to produce correct relic density and to satisfy DD cross
  section simultaneously. For illustrative purpose, let us choose a dark matter 
 mass with 500 GeV. From Fig.\ref{fig:sigmav-compare1}, we found that in order to satisfy the relic density, we need to have 
 a $\lambda^0_{\phi H} \sim$ 0.15 which can even be 0.02 in case with large 
 $\lambda_{\chi\phi}\sim 0.6$. Similarly we notice that for $m_{\textrm{DM}}=300$ GeV, $\lambda_{\phi H}^0$
  was 0.086 in order to produce correct relic density which however was excluded from direct search point of view. 
 This conclusion changes in presence of $\lambda_{\chi \phi}$ as we can see from Fig.\ref{correlation}, (left panel) 
 that $m_{\textrm{DM}} = 300$ GeV can produce correct relic density and evade the direct search limit with 
 smaller $\lambda_{\phi H}: 0.065 - 0.086$. 
 This is possible in presence of nonzero $\lambda_{\chi H}$ and small $\sin\theta$($\sim 0.2$ here) which redistribute the 
 previously obtained  value of $\lambda^0_{\phi H}$ into $\lambda_{\phi H}$  and $\lambda_{\chi \phi}$
 while simultaneously brings the direct search cross section less than the experimental limit due to its 
 association with $\sin\theta$ (see the definition of $\lambda_{1}$ and $\lambda_{1}$). 
 
\begin{figure}[htb!]
 \begin{center}
 \includegraphics[width=7.3cm,height=5cm]{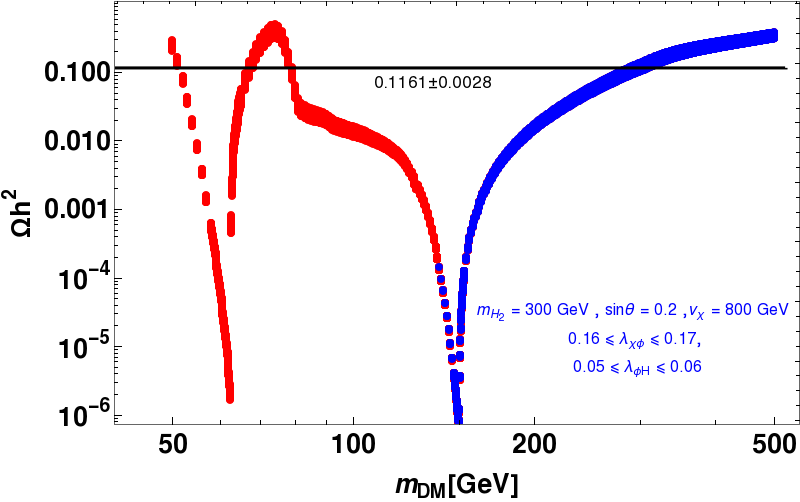}
 \includegraphics[width=7.3cm,height=5cm]{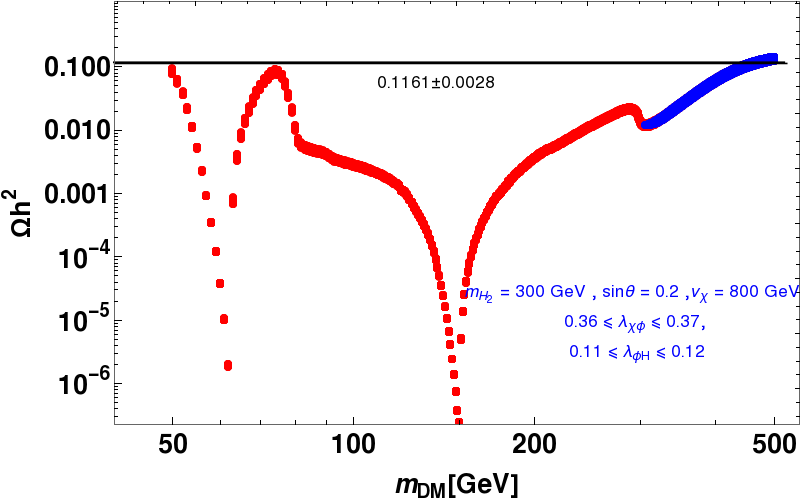}
 \end{center}
\caption{Relic density vs $m_{\textrm{DM}}$ plot in the combined set up of SM+DM+RH
 neutrinos and $\chi$ field for two different specified range of of $\lambda_{\phi H}$ and
  $\lambda_{\chi\phi}$ as mentioned within the inset of figures. Two resonances are clearly visible at $m_{\textrm{DM}}=m_{H_1}/2$ and 
  $m_{H_2}/2$ respectively
   Blue patch represents the favoured region by LUX 2016 direct detection cross 
  section limit
   whereas red patch is excluded by LUX 2016. }
  \label{fig:Omega_mDM}
 \end{figure}

  \begin{figure}[htb!]
 \begin{center}
 \includegraphics[width=8.5cm,height=5.8cm]{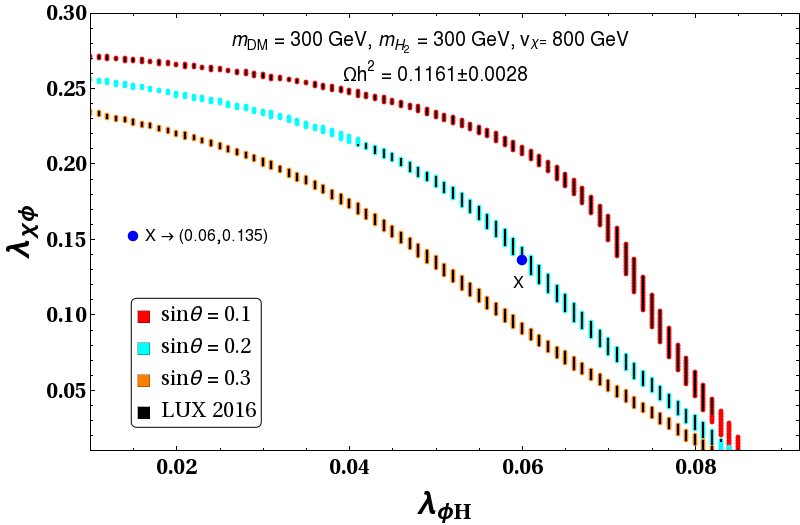}
 \end{center}
\caption{Allowed parameter space to satisfy correct relic abundance 
 in $\lambda_{\phi H}-\lambda_{\chi\phi}$ plane with different values of $\sin\theta$
  for $m_{\textrm{DM}}=300$ GeV. Other parameters $m_{H_2}=300$ GeV and $v_{\chi}=800$ GeV have been kept fixed. 
  The LUX 2016\cite{Akerib:2016vxi} allowed region are also accommodated (solid black region) in the figures.  The blue dot point denoted by  X in right
   panel
  will be used as a reference point for study on Higgs vacuum stability.}
  \label{fig:sinth}
 \end{figure}
 In Fig.\ref{fig:Omega_mDM} (left panel), we show the relic density versus $m_{\textrm{DM}}$ plot with our chosen set of parameters, 
 $\{m_{H_2}=300 \textrm{ GeV}$ ,$m_{H_1}=125.09$ GeV, $\tan\beta=0.307$, $\sin\theta=0.2\}$ 
 while varying $\lambda_{\chi H}$ and $\lambda_{\phi H}$ within $0.16\leq \lambda_{\chi\phi}\leq0.17$ and $0.05
 \leq\lambda_{\phi H}\leq 0.06$. Similarly in right panel, we provide the relic density vs $m_{\textrm{DM}}$ plot for a different 
 range of $\lambda_{\chi H}$ and $\lambda_{\phi H}$.
   We note that there are two resonance regions, one at $m_{H_1}/2$ for SM like Higgs and other at $m_{H_2}/2$
 with heavy Higgs\footnote{As expected, it would be always possible to satisfy the relic
  density and DD limits within this region.} mass at 300 GeV. In left panel for DM 
  heavier than 150 GeV, we find $m_{\textrm{DM}}\sim 300$ GeV 
  can correctly produce the relic density in the observed range and simultaneously evade the DD limit set by
  LUX 2016\cite{Akerib:2016vxi}. This result is consistent with the plot in Fig.\ref{correlation}. Similarly 
  $m_{\textrm{DM}}\sim 500$ GeV is in the acceptable range, which is in line with observation in Fig.\ref{correlation}.
   In the left panel of Fig.\ref{fig:Omega_mDM} we also have another region of DM mass$\sim 75$ GeV having correct 
   relic abundance however discarded by LUX 2016. 
   The region was not incorporated in top left panel of  Fig.\ref{correlation} as we have started with $m_{\textrm{DM}}$
   bigger than 150 GeV only. The possibility of having dark matter lighter than 150 GeV in the present scenario  
   will be discussed in the next subsection.   
 Since in obtaining the Fig.\ref{correlation}, we have fixed $\sin\theta, \tan\beta$ and $m_{H_2}$, below in
  Fig.\ref{fig:vchi} and \ref{fig:sinth}, we provide the 
 expected range of two couplings $\lambda_{\chi H}$ and $\lambda_{\phi H}$ when $\sin\theta, 
 \tan\beta$ are varied for dark matter mass $m_{\textrm{DM}} = 300$ GeV . 
 We find the variation is little sensitive with the 
 change of both $v_\chi$ and $\sin\theta$. As $v_\chi$ or $\sin\theta$ increases for $m_{\textrm{DM}}=300$ GeV, it requires less 
 $\lambda_{\chi\phi}$ for a particular $\lambda_{\phi H}$ to satisfy the relic density. We have also 
 applied the LUX 2016\cite{Akerib:2016vxi} DD cross section limit in those plots and are indicated by solid black patches.
  In  Fig.\ref{fig:sinth},
  one dark blue dot has been put on the $\sin\theta=0.2$ contour which will be used in study of Higgs vacuum stability
   as a reference point.
\begin{figure}[htb!]
 \begin{center}
 \includegraphics[width=7.3cm,height=6cm]{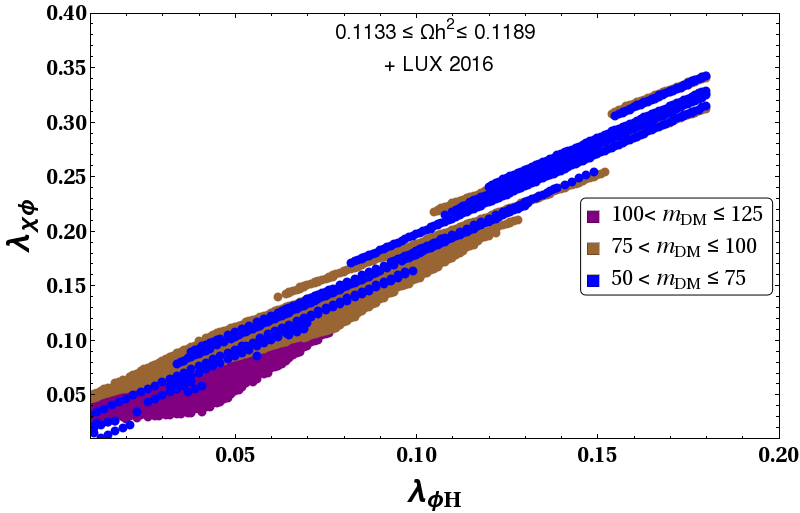}
 \includegraphics[width=7.3cm,height=6cm]{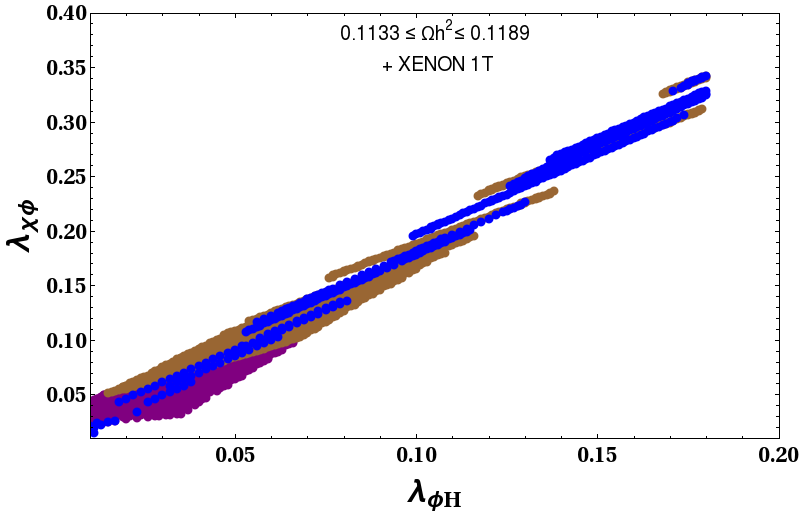}\\
 \end{center}
\caption{Relic density satisfied points in $\lambda_{\phi H}-\lambda_{\chi\phi}$ plane for $m_{\textrm{DM}}<150$ GeV
 with DD cross section consistent with [left panel] LUX 2016\cite{Akerib:2016vxi} and [right panel]
XENON 1T limit\cite{Aprile:2017iyp}. Benchmark points: $m_{H_2}=300$ GeV, $v_\chi=800$ GeV, 
$\sin\theta=0.2.$
  }
  \label{fig:lowmass}
 \end{figure}
 \begin{figure}[htb!]
 \begin{center}
 $$
 \includegraphics[width=8.5cm,height=5cm]{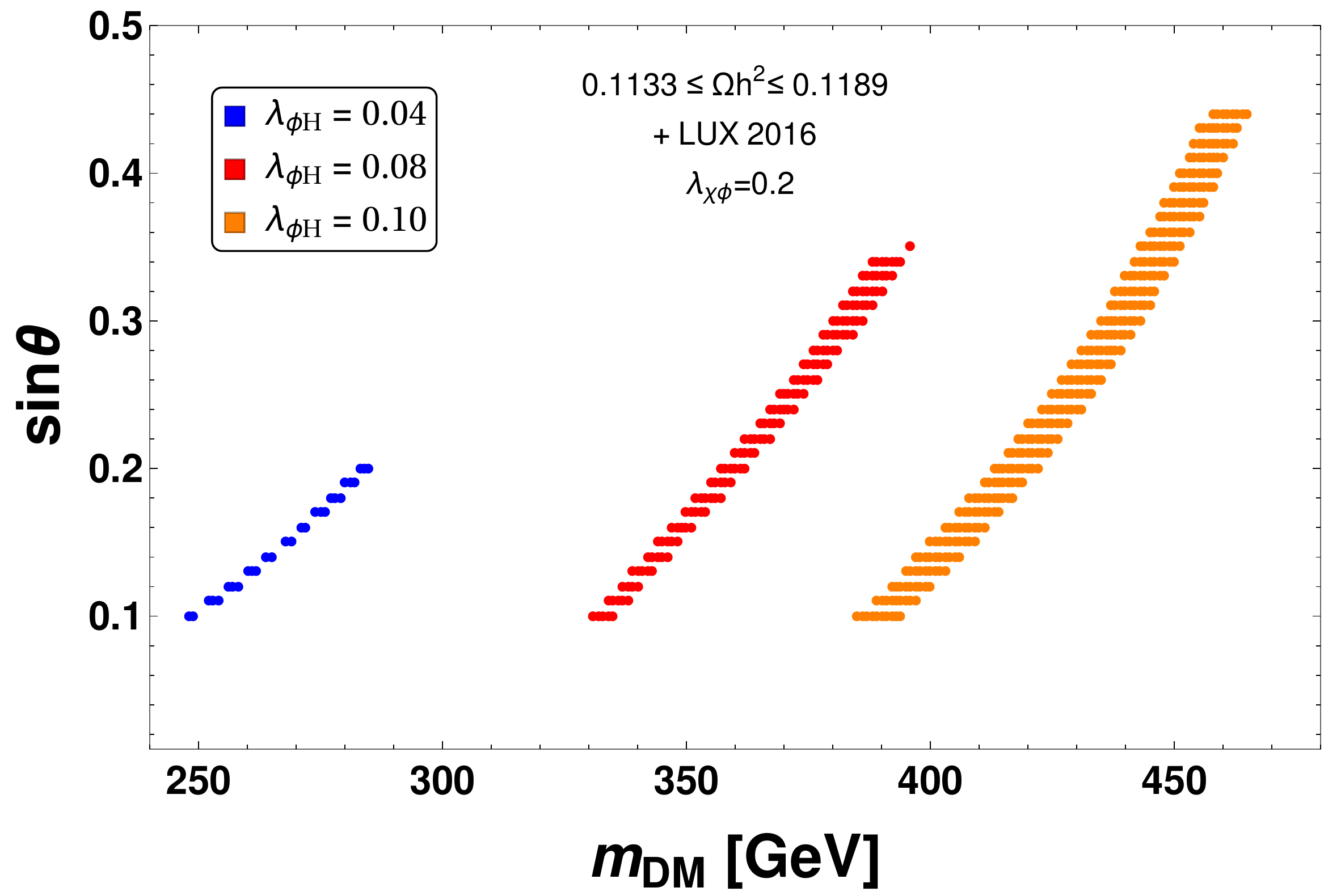}
 $$
 \end{center}
\caption{$m_{\textrm{DM}}$ vs $\sin\theta$ plot for a fixed $\lambda_{\chi\phi}$ and $\lambda_{\phi H}$ as mentioned
 in the figure to satisfy the correct relic abundance and direct detection cross section consistent with LUX 2016 limit. 
Values of other parameters: $m_{H_2}=300$ GeV, $\lambda_{\chi \phi}=0.2$ and $v_\chi=800$ GeV.}
  \label{fig:sinthmDM}
 \end{figure} 
\subsection{ DM mass in region R2: ($m_{\textrm{DM}}<150$ GeV)}
Here we briefly discuss the DM phenomenology in the low mass region $m_{\textrm{DM}}<\frac{m_{H_2}}{2}=150$ GeV.
  In this region, the decay process of heavy higgs to DM ($H_2\rightarrow \phi\phi$) will
   be active. For further low $m_{\textrm{DM}}<m_{H_1}/2\simeq 62.5$, both $H_2\rightarrow \phi\phi$ and $H_1\rightarrow \phi\phi$
   decay modes
   will be present.

We perform a scan over the $\lambda_{\phi H}-\lambda_{\chi\phi}$ region to find the 
   correct relic density satisfied parameter space with allowed direct detection cross section 
   from LUX 2016\cite{Akerib:2016vxi} and XENON 1T experiments\cite{Aprile:2017iyp}. The results are shown in
 Fig.\ref{fig:lowmass}, left and right panels where DD limits from LUX 2016 [left panel] and XENON 1T (preliminary) [right panel] 
 are considered separately. In doing these plots, we have considered different mass ranges as indicated by different colors. 
 The color codes are depicted within the inset of each figures. We note that the required $\lambda_{\chi\phi}$, $\lambda_{\phi H}$ 
values are almost in the similar range as obtained in Fig.\ref{correlation}. We also note that there exists a resonance region 
through $H_1$ near $m_{\textrm{DM}}\sim 63$ GeV, indicated by the blue patch. In this resonance region, the relic density becomes 
insensitive to the coupling and hence the blue patch is extended over the entire 
region of $\lambda_{\chi\phi}$, $\lambda_{\phi H}$
 in the Fig.\ref{fig:lowmass}.
 
Finally we attempt to estimate the $\sin\theta$ required to provide the 
correct amount of modification over the minimal version of a real singlet DM 
having interaction with SM Higgs only in order to revive the `below 500 GeV' 
DM into picture. In other words, the amount of $\sin\theta$ should be 
enough to satisfy correct relic abundance and DD cross section limits of 
LUX 2016\cite{Akerib:2016vxi} and XENON 1T\cite{Aprile:2017iyp} for this 
particular mass range. To do the analysis, we fix $\lambda_{\chi \phi}\sim0.2$ 
while three different values of $\lambda_{\phi H}$ at 0.04, 0.08 and 0.10 are 
considered for the study. We then provide the $\sin\theta$ versus $m_{\textrm{DM}}$ plot 
in Fig.\ref{fig:sinthmDM}
which is consistent with relic density and LUX 2016 limits. We infer that a sizable 
value of $\sin\theta$ is required for this. With $\lambda_{\phi H} = 0.1$, we have noted 
earlier from Fig. \ref{fig:sigmav-compare1} that it alone reproduces the desired relic density with a 330 GeV 
dark matter, although excluded by LUX 2016 limits. Now we observe from 
Fig. \ref{fig:sinthmDM} that in order to make this as a viable DM mass, we need to have a
$\sin\theta = \mathcal{O}$(0.1) with $\lambda_{\phi H} = 0.1$. Such a moderate value of $\sin\theta$ is
 compatible with LEP and LHC results. A larger value of $\sin\theta$ $\sim\mathcal{O}$(0.3)  with $\lambda_{\phi H} = 0.1$ 
can accommodate  DM mass around 440 GeV as seen from the Fig. \ref{fig:sinthmDM}. Similarly, 
we indicate that with $\lambda_{\phi H} = 0.08 [0.04]$ (for which DM mass $\sim$ $270$ GeV and $110$ GeV
satisfy the relic density as seen from Fig. \ref{fig:sigmav-compare1}),  $\sin\theta$ variation covers 
a range of DM mass $\sim$ 330-370 GeV [240-290 GeV] provided we restrict ourselves upto $\sin\theta = 0.3$.  

\section{Vacuum stability}\label{VSDMRH}

In this section, we will discuss how the EW vacuum stability can be achieved in our model. For clarification 
purpose and a comparative study of it, we first discuss how the presence of different
ingredients (three RH neutrinos, DM and extra scalar $\chi$) can affect the running of the Higgs quartic 
coupling when added one after other. We first comment on the inclusion of the RH neutrinos and investigate 
the running of $\lambda_H$. Then we study how the involvement of the scalar singlet DM field $\phi$ can alter 
the conclusion. Finally we discuss the result corresponding to our set-up,  $ i.e.$ including the $\chi$ field as well. 

In doing this analysis, the absolute stability of the Higgs vacuum is ensured by $\lambda_H(\mu)>0$ for 
 any energy scale $\mu$ where the EW minimum of the scalar potential is the global minimum.
 However there may exist another minimum which is deeper than the EW one. In that case we need to calculate 
 the tunneling probability of the EW vacuum to the second minimum. The Universe will be in metastable state provided
 the decay time of EW vacuum is longer than the age of the universe.
 The tunneling probability is given by\cite{Isidori:2001bm,Buttazzo:2013uya},
\begin{align}\label{Pro}
 \mathcal{P}=T_U^4\mu_B^4e^{-\frac{8\pi^2}{3|\lambda_H(\mu_B)|}},
\end{align}
 where $T_U$ is the age of the universe. $\mu_B$ is the scale at which probability is maximized, determined from $\beta_{\lambda_H}(\mu_B)=0$.
  Hence for metastable Universe requires\cite{Isidori:2001bm}
\begin{align}
 \lambda_H(\mu_B)>\frac{-0.065}{1-0.01\ln\Big(\frac{v}{\mu_B}\Big)},
\end{align}
 where $T_U\simeq 10^{14}$ yr is used. As noted in \cite{Buttazzo:2013uya}, for $\mu_B>M_P$, one can safely consider $\lambda_H(\mu_B)=\lambda_H(M_P$).

Before proceeding further, some discussion on the involvement of light neutrino mass in the context of
vacuum stbaility is pertinent here. As stated before, the light neutrino mass is generated through type-I seesaw for which 
three RH neutrinos are included in the set up. We now describe the strategy that we adopt here in order to study their
impact on RG evolution. For simplicity, the RH neutrino mass matrix $M_N$ is considered to be diagonal with degenerate entries,
{\it{i.e.}} $M_{i=1,2,3}=M_R$. As we will see, it is $\textrm{Tr}[Y_\nu^\dagger Y_\nu]$ which enters in the $\beta$ function
of the relevant couplings. In order to extract the information on $Y_\nu$, we employ the type-I mass
 formula $m_\nu=Y_{\nu}^T Y_{\nu} \frac{v^2}{2M_R}$. 
Naively one would expect that 
     large Yukawas are possible only with very large RH neutrino masses. For 
     example with $M_R \sim 10^{14}$ GeV, $Y_{\nu}$ comes out to be 0.3 in order to obtain $m_\nu\simeq 0.05$ eV. 
  Contrary to our naive expectation, 
     it can be shown 
     that even with smaller $M_R$ one can achieve large values of $\textrm{Tr}[Y_\nu^\dagger Y_\nu]$ once 
     a special flavor structure of $Y_{\nu}$ is considered\cite{Rodejohann:2012px}. Note that we aim to study the EW vacuum stability in presence of large
value of $\textrm{Tr}[Y_\nu^\dagger Y_\nu]$. For this purpose, we use the parametrization by \cite{Casas:2001sr} and write $Y_\nu$ as 
\begin{align}\label{casas}
    Y_\nu=\sqrt{2}\frac{\sqrt{M_R}}{v}  \mathcal{R}\sqrt{m_\nu^d} ~U^\dagger_{\textrm{PMNS}},
    \end{align}
where $m_\nu^d$ is the diagonal light neutrino mass matrix and $U_{\textrm{PMNS}}$ is the 
unitary matrix diagonalizing the neutrino mass matrix $m_{\nu}$ such that 
$m_{\nu} = U^*_{\textrm{PMNS}} m_\nu^d U^\dagger_{\textrm{PMNS}}$. Here  
$\mathcal{R}$ represents a complex orthogonal matrix which can be written as 
$\mathcal{R}=O\textrm{exp}(i\mathcal{A})$ with $O$ as real orthogonal and $\mathcal{A}$
as real antisymmetric matrices respectively. Hence one gets
 \begin{align}\label{trCa}     
\textrm{ Tr}[Y_\nu^\dagger Y_\nu] =  \frac{2M_R}{v^2}\textrm{Tr}\Big[\sqrt{m_\nu^d}e^{2i\mathcal{A}}\sqrt{m_\nu^d}\Big] . 
\end{align} 
Note that the real antisymmetric matrix 
     $\mathcal{A}$ does not appear in the seesaw expression for $m_{\nu}=\frac{Y_\nu^TY_\nu v^2}{2M_R}$.
 Therefore with any suitable choice 
     of $\mathcal{A}$, it would actually be possible to have sizeable Yukawas even with light $M_R$ and hence 
     this can affect the RG evolution of $\lambda_H$ significantly. As an example, 
     let us consider magnitude of all the entries of $\mathcal{A}$ to be equal, say $a$ with all diagonal entries as zero. 
     Then with $M_R$ = 1 TeV, 
     $\textrm{Tr}[Y_\nu^\dagger Y_\nu]$ can be as large as 1 with $a =8.1 $\cite{Casas:2001sr,Guo:2006qa}.
 Below we specify the details of Higgs vacuum stability
in presence of RH neutrinos only.

\subsection{Higgs vacuum stability with right-handed neutrinos}\label{SMRH}
 In presence of the RH neutrino Yukawa coupling $Y_{\nu}$, the renormalization group 
(RG) equation of SM couplings will be modified\cite{Pirogov:1998tj}. 

Below we present the one loop beta functions of Higgs quartic coupling $\lambda_H$, top quark Yukawa coupling 
$y_t$ and neutrino Yukawa coupling $Y_\nu$, 
\begin{align}
&\frac{d \lambda_H}{d\textrm{ln}\mu}=\frac{1}{16\pi^2}\{\beta_{\lambda_H}^{SM}+\beta_{\lambda_H}^\textrm{I}\}\textrm{  with  }
\beta_{\lambda_H}^\textrm{I}= 4\lambda_H\textrm{Tr}[Y_\nu^\dagger Y_\nu]-2 \textrm{Tr}[(Y_\nu^\dagger Y_\nu)^2]\,,\label{RL1}\\
&\frac{d y_t}{d\textrm{ln}\mu}=\frac{1}{16\pi^2}\{\beta_{y_t}^{SM}+\beta_{y_t}^\textrm{I}\} \textrm{  with  }\beta_{y_t}^\textrm{I}=
\textrm{Tr}[Y_\nu^\dagger Y_\nu]y_t,\label{RL2}\\
&\frac{d\textrm{Tr}[Y_\nu^\dagger Y_\nu] }{d\textrm{ln}\mu}=\frac{1}{16\pi^2}\beta_{\textrm{Tr}[Y_\nu^\dagger Y_\nu]}^\textrm{I}=\frac{1}{16\pi^2}
\Big\{(6 y_t^2+2 \textrm{Tr}[Y_\nu^\dagger Y_\nu]-\frac{3}{2}g_1^2-\frac{9}{2} g_2^2)\textrm{Tr}[Y_\nu^\dagger Y_\nu]+
3\textrm{Tr}[(Y_\nu^\dagger Y_\nu)^2]\Big\},
\label{eq:RGnu}
\end{align}  
where $\beta_{\lambda_H}^{SM}$ and $\beta_{y_t}^{SM}$ represent the $\beta$ functions of $\lambda_H$
 and $y_t$ respectively in SM. The $Y_{\nu}$ dependence is to be evaluated in accordance with the type-I seesaw expression, 
$m_\nu=Y_{\nu}^T Y_{\nu} \frac{v^2}{M_R}$. 
\begin{figure}[h]
 \begin{center}
\includegraphics[width=7cm, height=5cm]{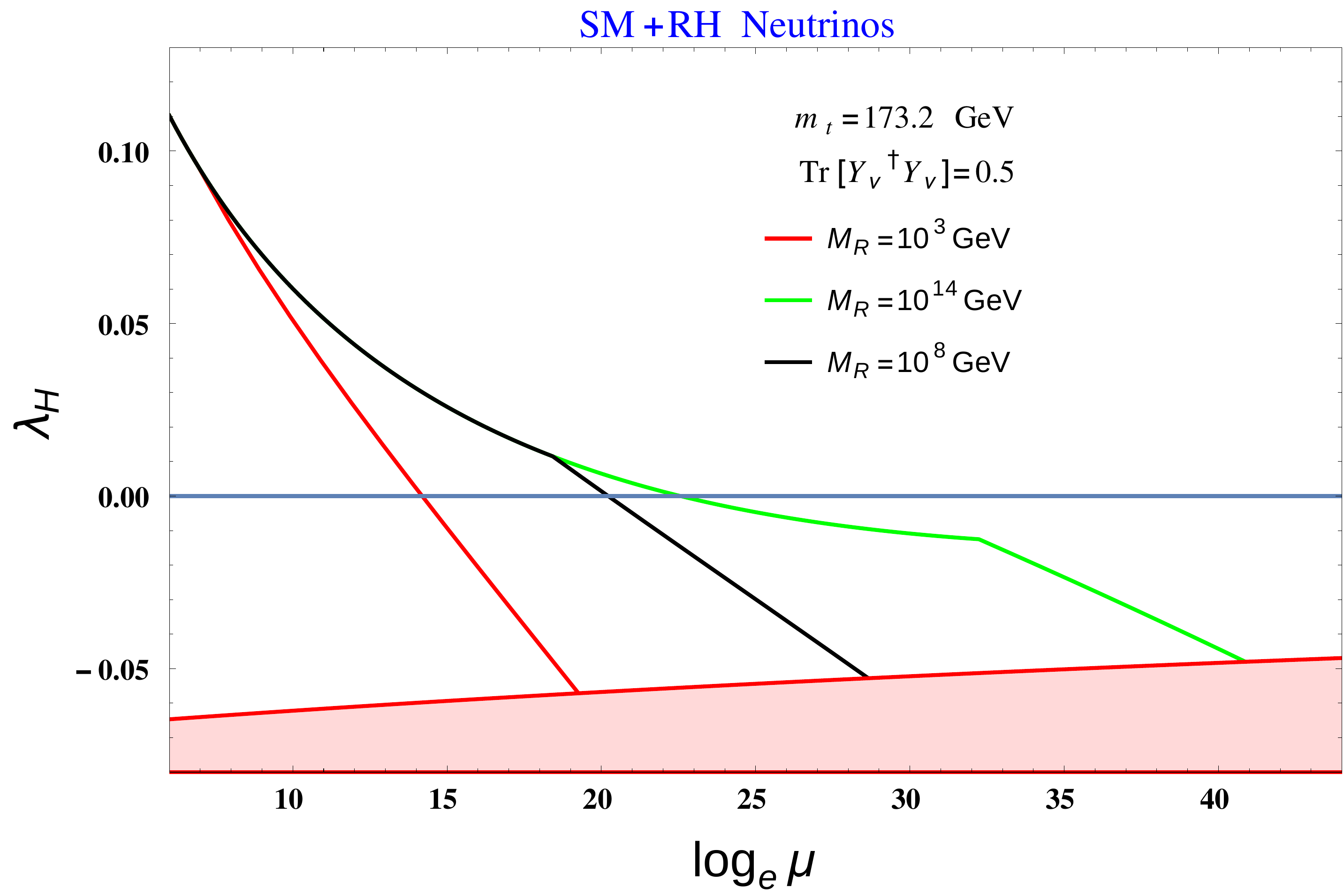}
  \includegraphics[width=7cm, height=5cm]{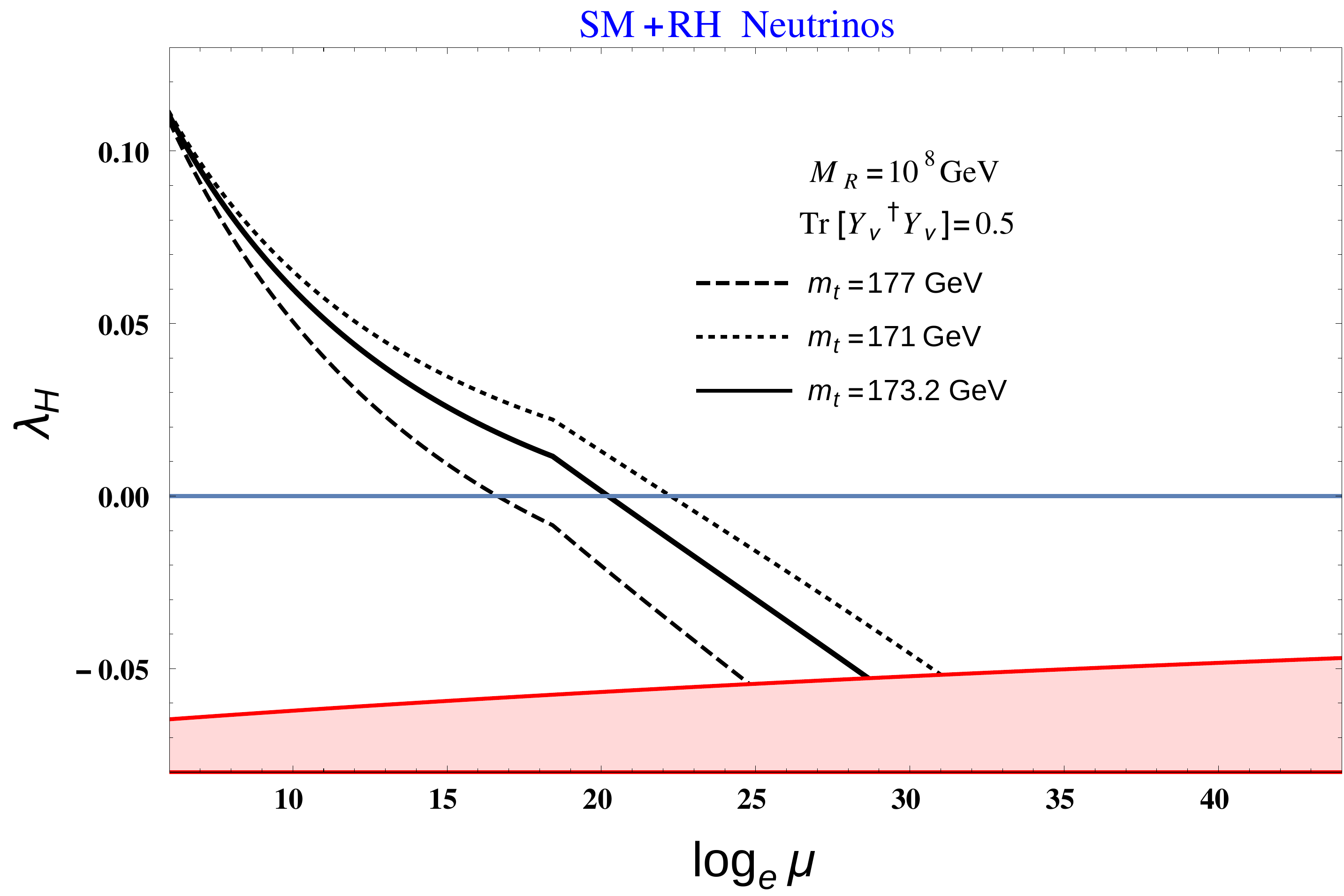}
 \end{center}
\caption{RG running of $\lambda_H$ with energy scale $\mu$ in SM + RH neutrinos;[Left panel]: different 
RH neutrino mass scales $M_R$ are considered with fixed $m_t =173.2 $ GeV, [Right panel]: different
  top masses are considered with $M_R = 10^8$ GeV.}
  \label{fig:SMrunning}
 \end{figure}
Also with large $a$ (elements of $\mathcal{A}$), it is found \cite{Rodejohann:2012px} 
     that Tr$[(Y_\nu^\dagger Y_\nu)^2]\simeq \textrm{Tr}[Y_\nu^\dagger Y_\nu]^2$ and we will be using this 
     approximated relation in obtaining the running of the couplings through Eqs.(\ref{RL1},\ref{RL2},\ref{eq:RGnu}).
 Here we have used the best fit values of neutrino oscillation parameters for normal
  hierarchy\cite{Esteban:2016qun,deSalas:2017kay}. We have also
 considered the mass of lightest neutrino to be zero.
 
\begin{figure}[h]
 \begin{center}
  \includegraphics[width=7cm, height=5cm]{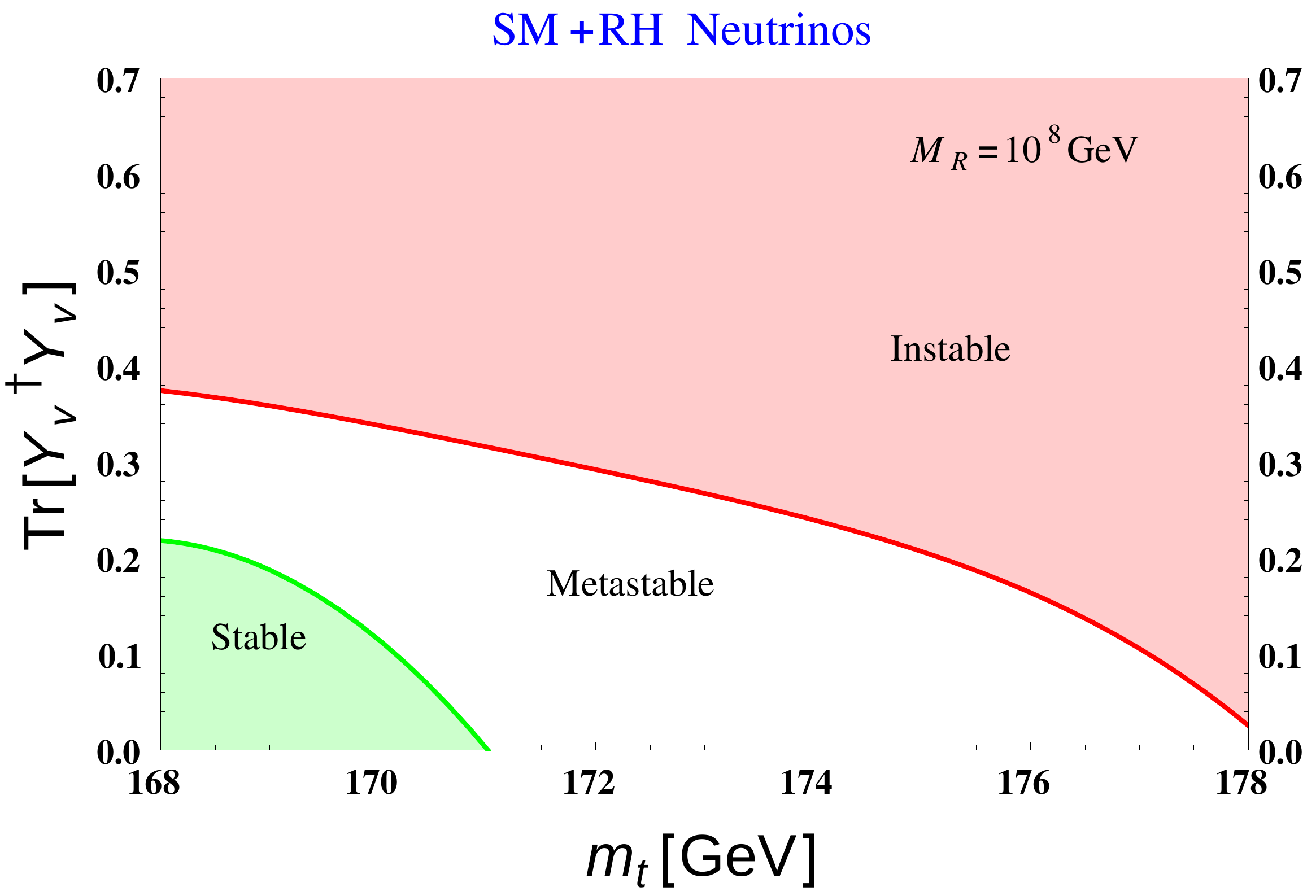}
 \end{center}
\caption{Region plot for $m_t$-$\textrm{Tr}[Y_\nu^\dagger Y_\nu]$ in
 the SM, extended with RH neutrinos having degenerate mass $M_R=10^8$ GeV. The plane is divided into three
  categories (i) absolute stability, (ii) metastability and (iii) instability.}
  \label{fig:SMCont}
 \end{figure}
 Note that just like the top quark Yukawa coupling, the neutrino Dirac Yukawa is having a similar 
impact on the Higgs quartic coupling, in particular with large $Y_{\nu}$. Also the top quark Yukawa would have a 
contribution dependent on $Y_{\nu}$. This has been studied in several works\cite{Chen:2012faa,Rodejohann:2012px,Rose:2015fua,Bambhaniya:2016rbb,Lindner:2015qva,Chakrabortty:2012np,Datta:2013mta,Coriano:2014mpa,Ng:2015eia,Bonilla:2015kna}. We summarize here the results with some benchmark values of RH neutrino masses. These will be useful for a 
comparative study with the results specific to our model. 
In Fig.\ref{fig:SMrunning} (left panel), we have plotted running of the Higgs quartic coupling $\lambda_H$ 
against energy scale $\mu$ till $M_P$ for different choices of 
$M_R = 10^3, 10^8$ and $10^{14}$ GeV with $\textrm{Tr}[Y_\nu^\dagger Y_\nu] = 0.5$ denoted by 
red, black and green solid lines respectively. The pink shaded portion represents the instability region 
given by the inequality \cite{Isidori:2001bm} $\lambda_H\leq {-0.065}/[1-0.01\textrm{ln}\Big(\frac{v}{\mu}\Big)]$.
As expected, we find that the Higgs quartic coupling enters into the instability region well before the Planck scale.

\begin{table}[h]
\begin{center}
\begin{tabular}{|c | c | c | c | c | c |}
\hline
Scale & $y_t$ & $g_1$ & $g_2$ & $g_3$ & $\lambda_H$\\
\hline
$\mu=m_t $ & $0.93610$ & $0.357606$ & $0.648216$ & $1.16655$ & $0.125932$\\
\hline
\end{tabular}
\end{center}
\caption{Values of  the relevant SM couplings (top-quark Yukawa $y_t$, gauge couplings $g_i$ and $\lambda_H$) at
 energy scale $\mu=m_t=173.2$ GeV with $m_h=125.09$ GeV and $\alpha_S(m_Z)=0.1184$.}
\label{tab:ini}
\end{table}  
In the right panel of Fig.\ref{fig:SMrunning}, the effect of choosing different $m_t$ within the 
present $2\sigma$ uncertainty is shown for a fixed $M_R = 10^8$ GeV. The black solid, dashed 
and dotted lines represent the $\lambda_H$ running with $m_t$ as 173.2 GeV, 177 GeV and 171 GeV 
respectively.
In doing this analysis, we fix the initial values of all SM couplings \cite{Buttazzo:2013uya} as given in 
Table \ref{tab:ini} at an energy scale $\mu=m_t$. Here we consider $m_h=125.09$ GeV, $m_t=173.2$ 
GeV and $\alpha_s=0.1184$. 
In Fig.\ref{fig:SMCont}, we have shown a region plot for $\textrm{Tr}[Y_\nu^\dagger Y_\nu]$ and
 $m_t$ with fixed $M_R$ at $10^{8}$ GeV in terms of stability ($\lambda_H$ remains positive all the way upto $M_P$),
  metastability and
 instability of the EW vacuum of the SM.
The top quark mass is varied between 168 GeV to 178 GeV. The region in which EW vacuum is stable is indicated by green 
and the metastable region is indicated by white patches. The instability region is denoted with pink shaded part. It can be 
noted that the result coincides with the one obtained in \cite{Lindner:2015qva}. We aim to discuss the change obtained 
over this diagram in the context of our model. 
\subsection{Higgs vacuum stability from Higgs Portal DM and RH neutrinos}\label{SMDMRH}
 Here we discuss the vacuum stability scenario in presence of both the scalar DM ($\phi$) and 
 three RH neutrinos ($N$). In that case, effective scalar potential becomes $V_{\textrm{I}} + V_H$ only. 
 Note that the DM phenomenology is essentially unaffected from the inclusion of the heavy 
 RH neutrinos with the assumption $M_R\gg m_{\textrm{DM}}$. 
 On the other hand combining Eq.(\ref{RL1},\ref{RL2},\ref{eq:RGnu}),
 we obtain the corresponding beta functions for the couplings as provided below;
\begin{align}
\frac{d\lambda_H}{dt}&=\frac{1}{16\pi^2}\{\beta_{\lambda_H}^{SM}+\beta_{\lambda_H}^{\textrm{I}}+\beta_{\lambda_H}^{\textrm{II}}\}\textrm{ where }
 \beta_{\lambda_H}^{\textrm{II}}=\frac{\lambda_{\phi H}^2}{2},\\
\frac{d\lambda_{\phi H}}{dt}&=\frac{1}{16\pi^2}\beta_{\lambda_{\phi H}}^\textrm{I}=\frac{1}{16\pi^2}
\Big\{12\lambda_H\lambda_{\phi H}+\lambda_\phi\lambda_{\phi H}+4\lambda_{\phi H}^2+6 y_t^2
 \lambda_{\phi H}-\frac{3}{2}g_1^2\lambda_{\phi H}-\frac{9}{2}g_2^2\lambda_{\chi H}+2\textrm{Tr}[Y_
\nu^\dagger Y_\nu]\lambda_{\phi H}\Big\},\\
\frac{d\lambda_{\phi}}{dt}&=\frac{1}{16\pi^2}\beta_{\lambda_{\phi}}^\textrm{I}=\frac{1}{16\pi^2}\Big\{3\lambda_\phi^2+12\lambda_{\phi H}^2\Big\}.\nonumber\\
\end{align}
 From the additional term  $\beta_{\lambda_{H}}^{\textrm{II}}$, we expect that the involvement of 
 DM would affect the EW vacuum stability in a positive way ($i.e.$ pushing the vacuum more toward 
 the stability) as shown in \cite{Haba:2013lga,Khan:2014kba,Khoze:2014xha,Gonderinger:2009jp,Gonderinger:2012rd,Chao:2012mx}
  whereas we noted in the previous subsection that the Yukawa coupling 
 (if sizable) has a negative impact on it.

\begin{figure}[h]
 \begin{center}
 \includegraphics[width=7cm, height=5cm]{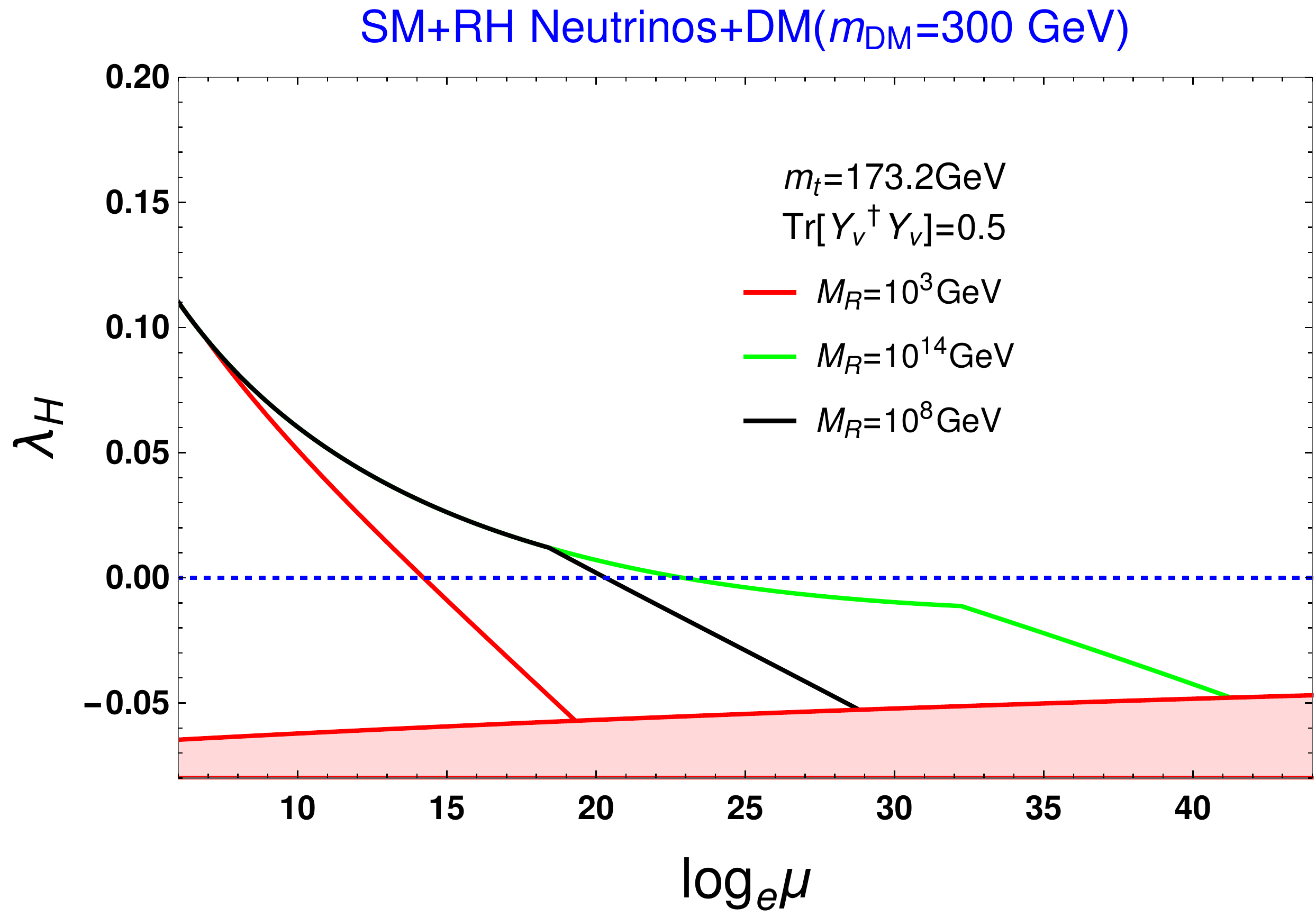}
 \includegraphics[width=7cm, height=5cm]{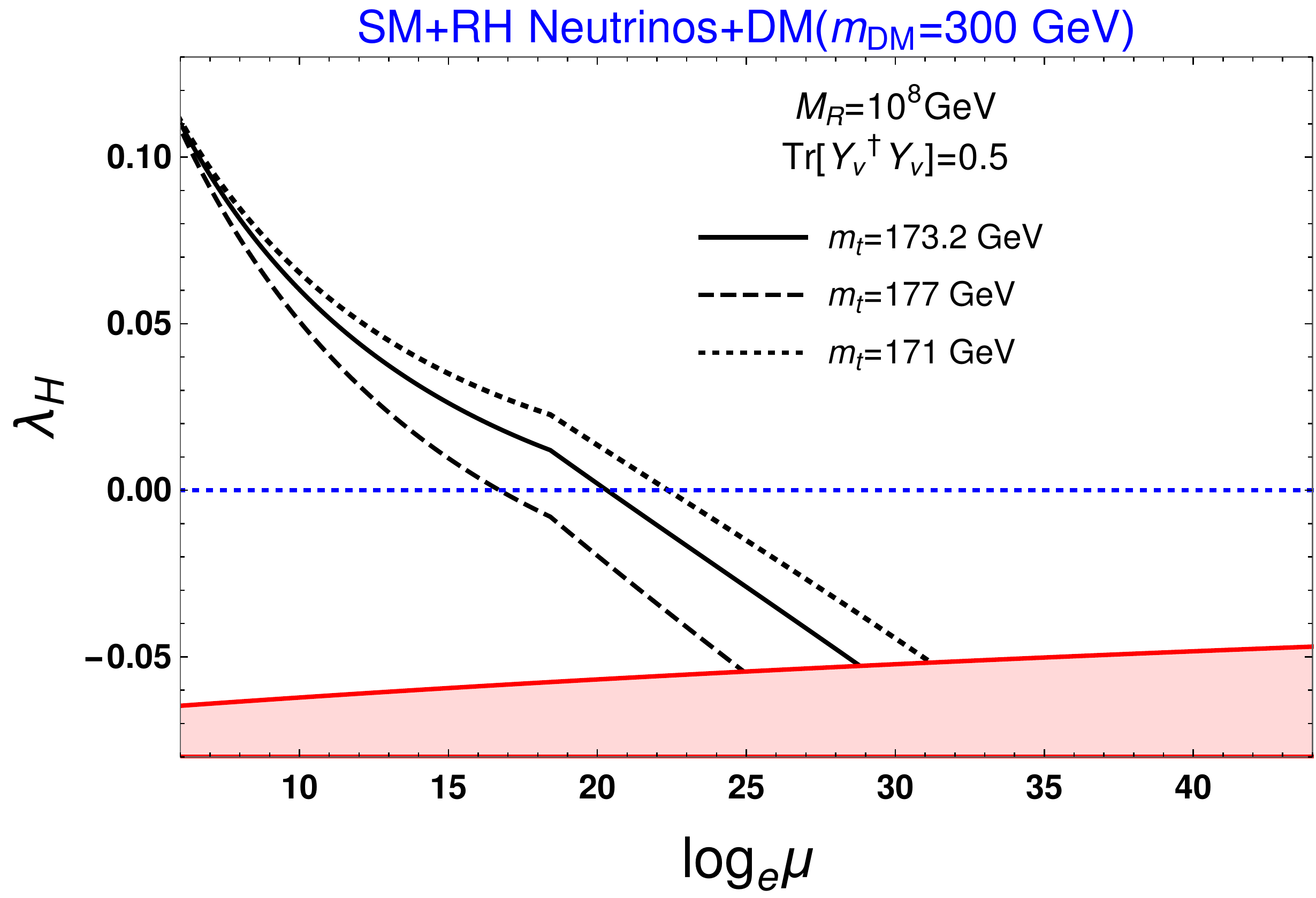}\\
 \includegraphics[width=7cm, height=5cm]{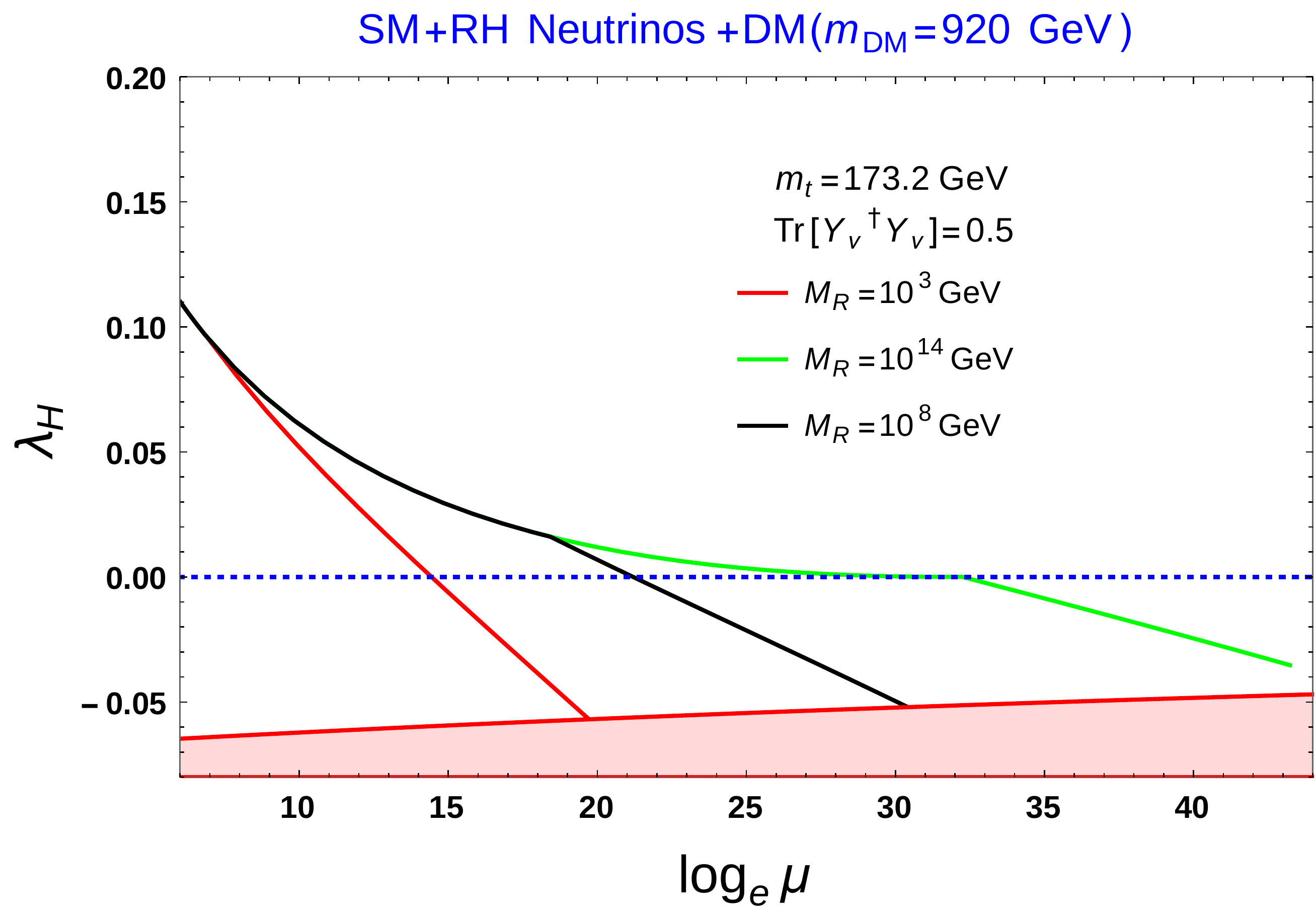}
 \includegraphics[width=7cm, height=5cm]{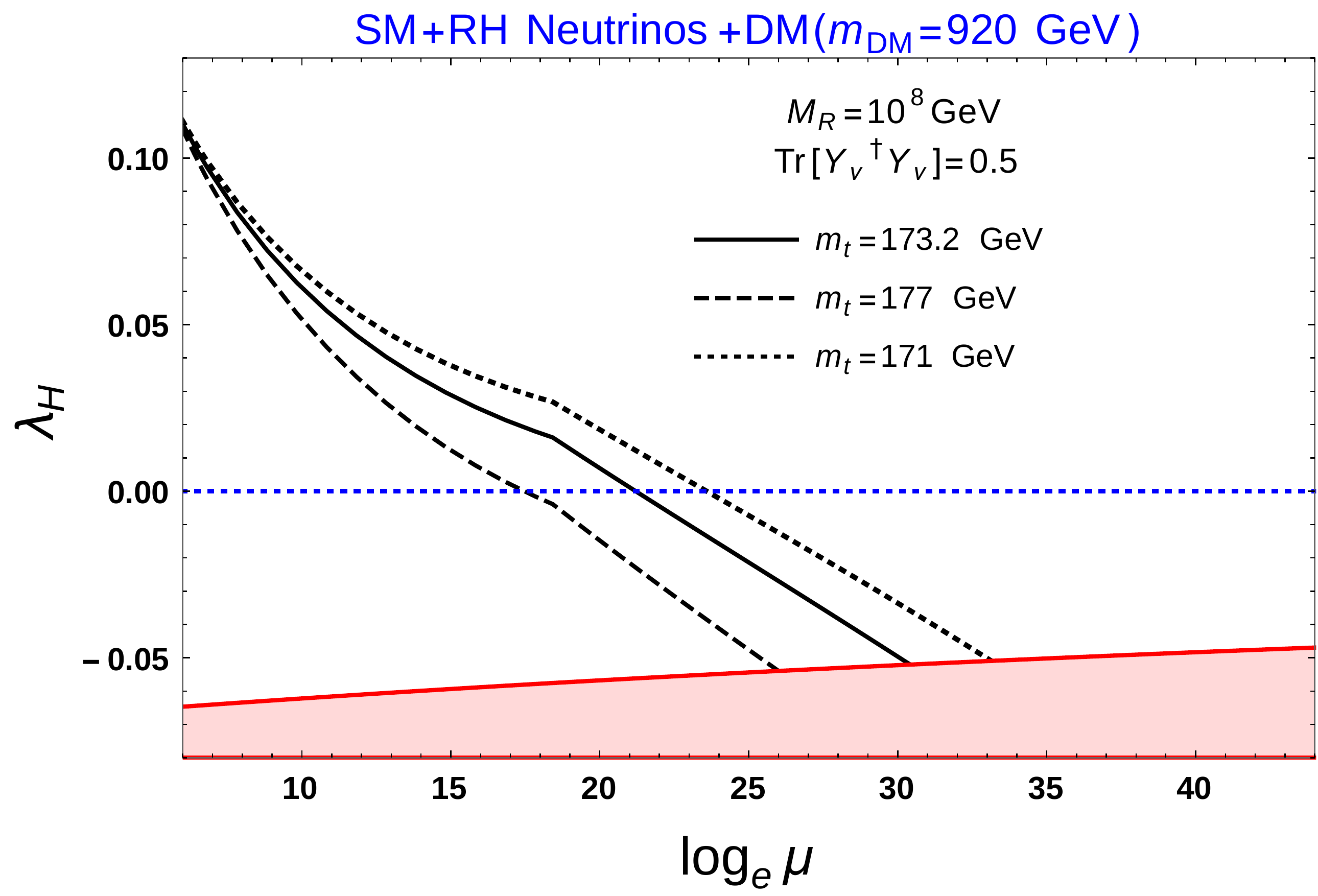}
 \end{center}
\caption{RG evolution of $\lambda_H$ with energy scale $\mu$ with SM+DM+RH Neutrinos with $\lambda_\phi=0.7$, $m_H=125.09$ GeV and $\alpha_S(m_Z)=0.1184$.: 
(a) [top panels]: $m_{\textrm{DM}}=300$ GeV, and (b) 
[bottom panels]: $m_{\textrm{DM}}=920$ GeV. In left panels $m_t$ is fixed at 173.2 GeV and plots are there with different 
$M_R$ while in right panel $M_R$ is fixed at $10^8$ GeV and different $m_t$ values are considered.}
\label{VSDMrunning}
\end{figure}

\begin{figure}[h]
 \begin{center}
 \includegraphics[width=7cm, height=5cm]{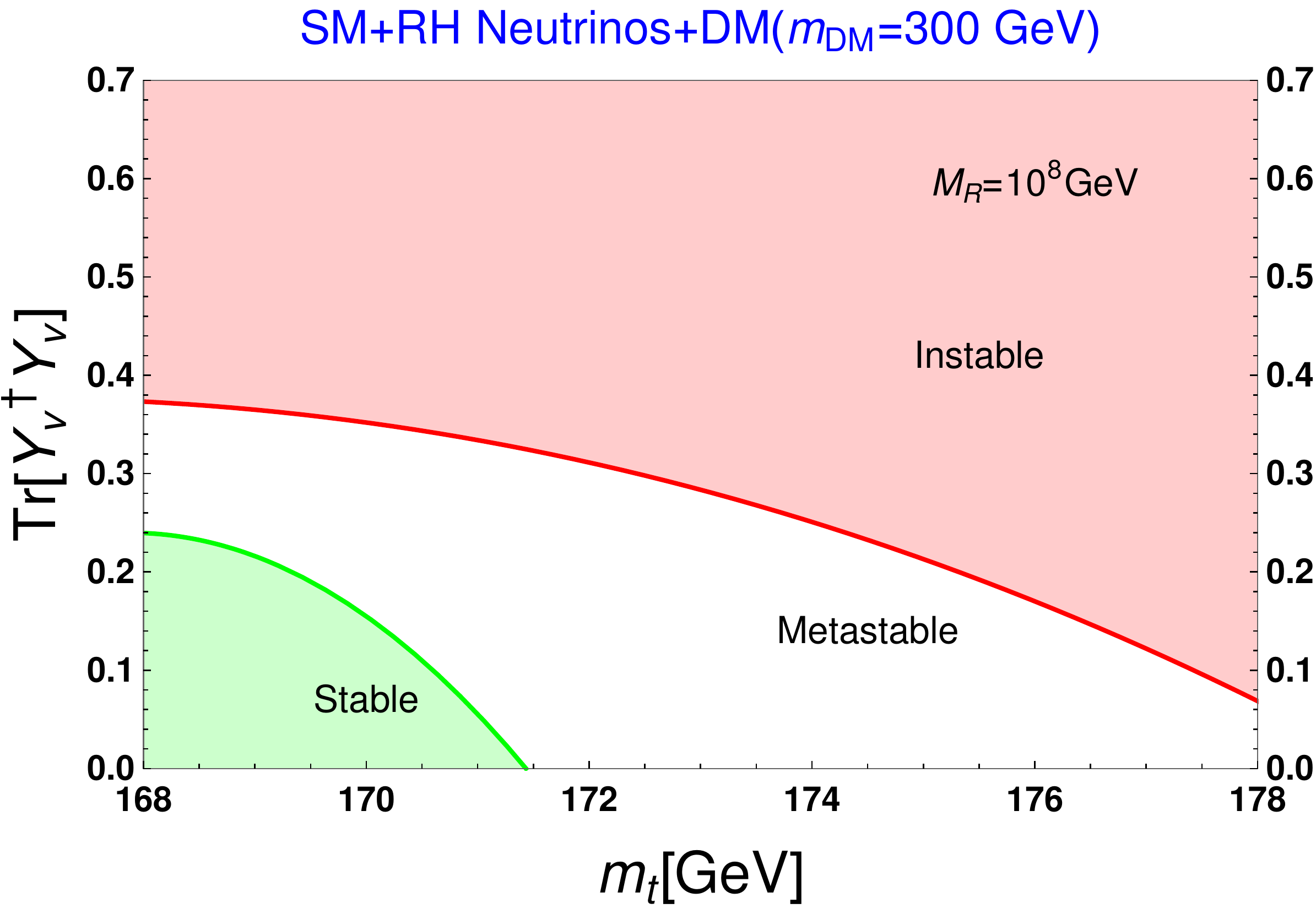}
 \includegraphics[width=7cm, height=5cm]{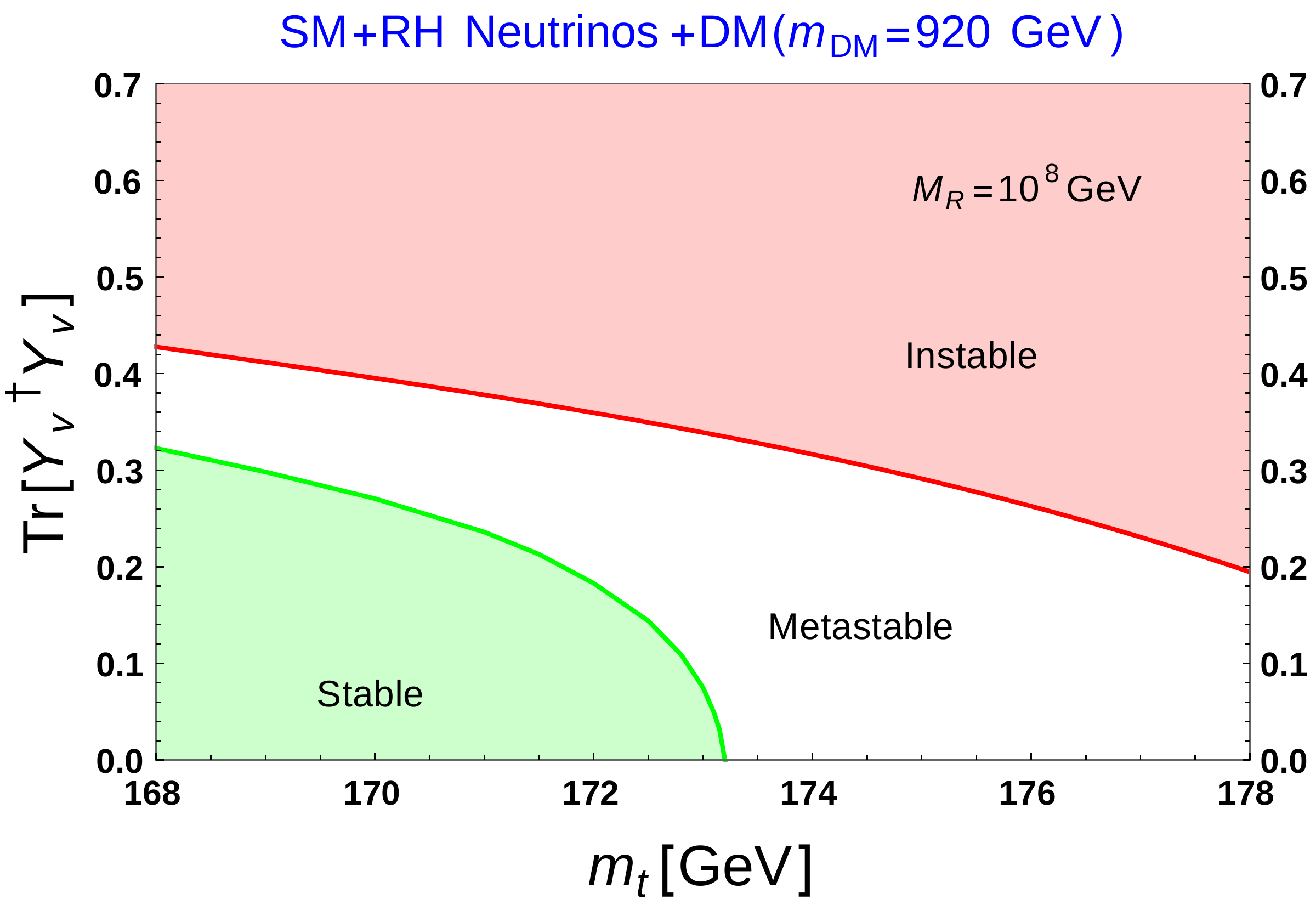}
 \end{center}
\caption{Regions of Stability, metastability and instability in SM+DM+RH neutrinos case
 in the $\textrm{Tr}[Y_{\nu}^\dagger Y_\nu ]$ -$m_t$ 
plane for $m_{\textrm{DM}}=300$ GeV (left panel) and $920$ GeV (right panel).
 We consider $\lambda_\phi=0.7$, $m_H=125.09$ GeV and $\alpha_S(m_Z)=0.1184$
 for both the figures. }
  \label{VSDMcontour}
 \end{figure}

 The interplay between the neutrino Yukawa coupling and Higgs portal coupling with DM is shown in Fig. \ref{VSDMrunning}, left 
and right panels (top and bottom). 
For the purpose of comparison, we have kept the same set of choices of parameters as in Fig.\ref{fig:SMrunning}, (left and right panels ).
 For the top panels, we consider mass of the dark matter to be $m_{\textrm{DM}}=300$ GeV
  and for the bottom set, $m_{\textrm{DM}}=920$ GeV is taken.  
The choice of $m_{\textrm{DM}}$ could in turn fix the $\lambda_{\phi H}$ coupling from the relic density plot of Fig.\ref{fig:sigmav-compare1}. For
  example with $m_{\textrm{DM}}=300$ GeV $\lambda_{\phi H}$ is 0.075 and for $m_{\textrm{DM}}=920$ GeV, $\lambda_{\phi H}$
   is given by 0.286 value. It is evident that the presence of Higgs portal 
 coupling only has a mild effect as compared to the  impact created by the neutrino Yukawa coupling.
 Finally in Fig.\ref{VSDMcontour} we provide the region plot in Tr[$Y^{\dagger}_{\nu} Y_{\nu}$] - $m_t$ plane where the 
 stable and instable regions are indicated by green and pink patches. This plot while compared with 
 Fig.\ref{fig:SMCont}, indicates that there is no such noticeable improvement except the mild enhancement of the 
 metastable region due to the involvement of singlet scalar (DM) with Higgs portal coupling. With an aim 
 to accommodate both the massive neutrinos and a relatively light dark matter ($<$ 500 GeV), we move on 
 to the next section where the $\chi$ field is included.

\subsection{EW vacuum stability in extended Higgs portal DM and RH neutrinos}

Turning into the discussion on vacuum stability in our framework of extended Higgs portal having 
three RH neutrinos, DM and the $\chi$ fields, we first put together the relevant RG equations 
(for $\mu > m_{\textrm{DM}}, m_{H_2}$) 
 as given by, 
\begin{align}
 \frac{d\lambda_H}{dt}&=\frac{1}{16\pi^2}\{\beta_{\lambda_H}^{SM}+\beta_{\lambda_H}^{\textrm{I}}+\beta_{\lambda_H}^{\textrm{II}}+
 \frac{\lambda_{\chi H}^2}{2}\},\\
 \frac{d\lambda_{\phi H}}{dt}&=\frac{1}{16\pi^2}\{\beta_{\lambda_{\phi H}}^{\textrm{I}}+\lambda_{\chi \phi}\lambda_{\chi H}\},\\
 \frac{d\lambda_{\phi}}{dt}&=\frac{1}{16\pi^2}\{\beta_{\lambda_{\phi }}^{\textrm{I}}+3\lambda_{\chi\phi}^2\},\\
 \frac{d\lambda_{\chi H}}{dt}&=\frac{1}{16\pi^2}\Big\{12\lambda_H\lambda_{\chi H}+\lambda_\chi\lambda_{\chi H}+4\lambda_{\chi H}^2+6 y_t^2
 \lambda_{\chi H}-\frac{3}{2}g_1^2\lambda_{\chi H}-\frac{9}{2}g_2^2\lambda_{\chi H}+\lambda_{\chi \phi}\lambda_{\phi H}+2\textrm{Tr}[Y_
\nu^\dagger Y_\nu]\lambda_{\chi H}\Big\},\nonumber\\
 \frac{d\lambda_{\chi}}{dt}&=\frac{1}{16\pi^2}\Big\{3\lambda_\chi^2+12\lambda_{\chi H}^2+3\lambda_{\chi\phi}^2\Big\},\nonumber\\
 \frac{d\lambda_{\chi\phi}}{dt}&=\frac{1}{16\pi^2}\Big\{4\lambda_{\chi\phi}^2+\lambda_{\chi\phi}(\lambda_\phi+\lambda_\chi)+4\lambda_{\phi H}\lambda_{\chi H}\Big\}\label{RGlambdachiphi}.
\end{align}
We note that the couplings $\lambda_{\chi \phi}, \lambda_{\phi H}$ and $\lambda_{\chi H}$ which played important role 
in DM phenomenology, are involved in the running of couplings as well. From the discussion of the DM section, we 
have estimated these parameters in a range so as to satisfy the appropriate relic density and be within the direct search 
limits for a specific choice of other parameters at their reference values: $m_{H_2}=300$ GeV and $v_\chi=800$ GeV,
 $\sin\theta=0.2$ (henceforth we describe 
this set as $A$). In particular an estimate for $\lambda_{\chi \phi}, \lambda_{\phi H}$ are obtained from 
Fig.\ref{correlation} (for 150 GeV$< m_{\textrm{DM}} <$ 500 GeV) and from Fig. \ref{fig:lowmass}  (for $m_{\textrm{DM}} <$ 150 GeV) having 
different choices of $m_{\textrm{DM}}$ and $\sin\theta$. The parameter $\lambda_{\chi H}$ dependence is mostly realized 
through $\sin\theta$ following Eq.(\ref{lambdaChiH}), where $m_{H_2}, \tan\beta$ are fixed from set $A$. 
This $\sin\theta$
is the most crucial parameter which control both the DM phenomenology and the vacuum stability. We have already 
seen that it allows the scalar singlet DM to be viable for the low mass window by relaxing $\lambda_{\phi H}$ from 
its sole role in case of single scalar singlet DM. On the other hand, a non-zero $\sin\theta$ provides a 
positive shift (it is effectively the threshold effect in the small $\theta$ limit as seen from Eq. (\ref{lambdaH})) to the Higgs quartic 
coupling and hence guides the $\lambda_H$ toward more stability. Hence $\sin\theta$ would be a crucial parameter 
in this study.  Note that the RH neutrinos being relatively heavy as compared to the DM, neutrino Yukawa coupling 
does not play much role in DM phenomenology.

\begin{figure}[htb!]
 \begin{center}
 \includegraphics[width=7cm,height=5cm]{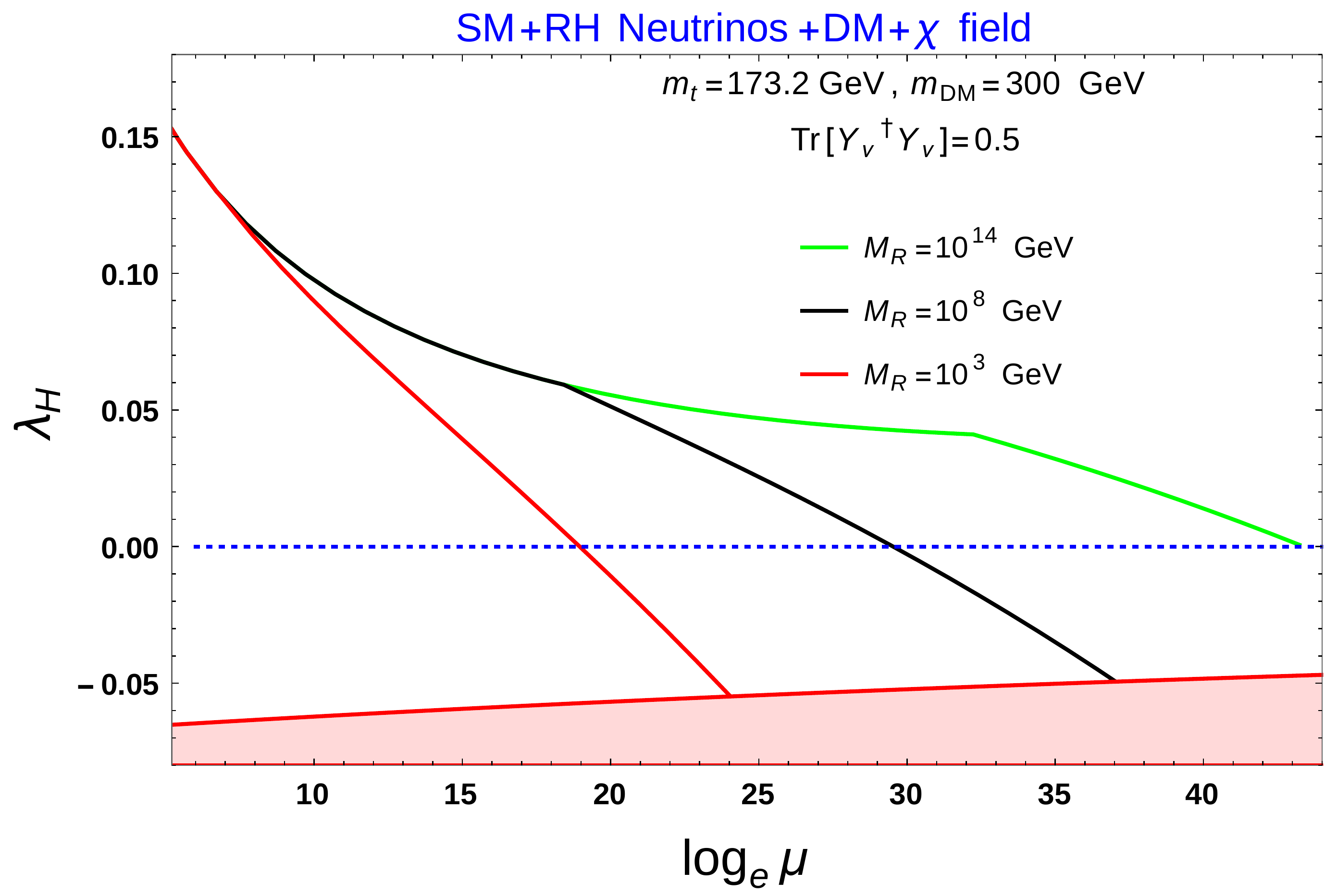}
 \includegraphics[width=7cm,height=5cm]{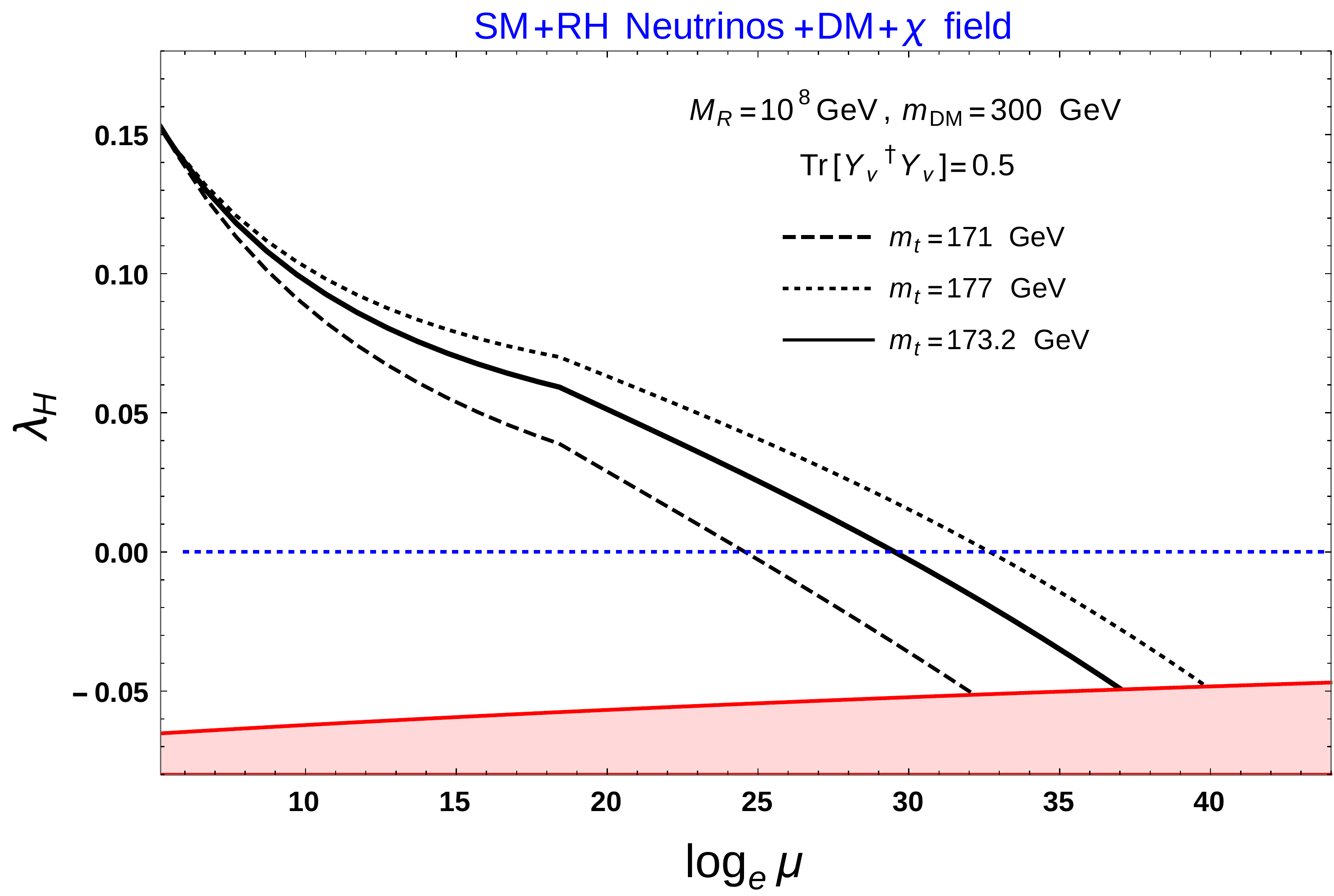}
 \end{center}
\caption{RG running of $\lambda_H$ vs $\mu$ in the combined scenario of SM+RH Neutrinos+DM+$\chi$ field with $m_{\textrm{DM}}=300$ GeV, 
$\sin\theta=0.2$ and $m_{H_2}=300$ GeV.
In [left pannel] $m_t$ ($\sim 173.2$ GeV) is kept fixed, $M_R$ is varied, and in [right panel]
 $M_R$  ($\sim 10^{8}$ GeV)  is fixed, $m_t$ has been varied.
 Point X ($\lambda_{\phi H}=0.06$, $\lambda_{\chi\phi}=0.135$)
 from Fig.\ref{fig:sinth} and $\lambda_{\phi}= 0.7$ have been used as benchmark points. }
  \label{fig:running3}
 \end{figure}
Assuming the validity of this extended SM (with three RH neutrinos and two singlets, $\phi, \chi$) upto the Planck scale,
\begin{figure}[htb!]
 \begin{center}
 \includegraphics[width=7cm,height=5cm]{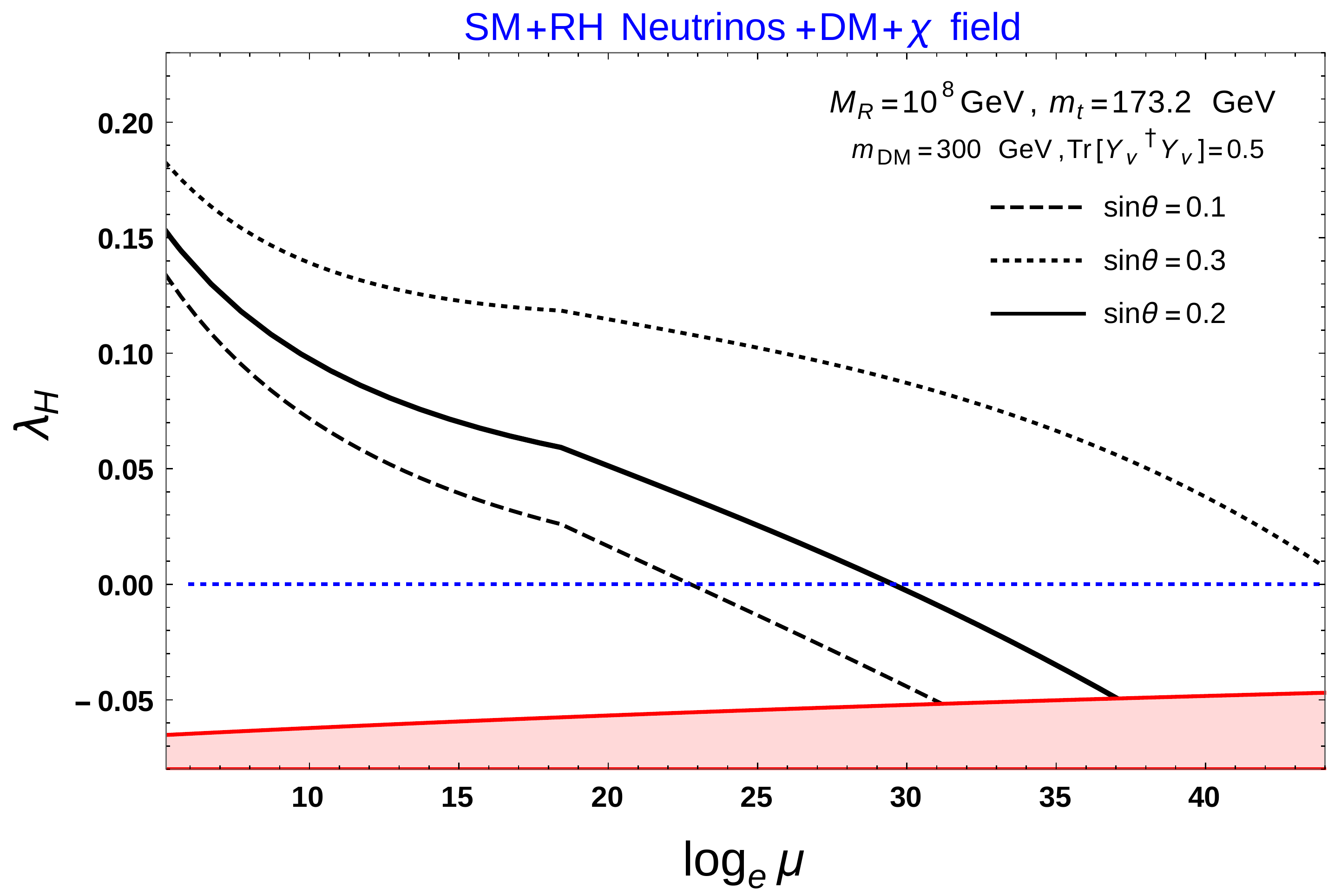}
 \includegraphics[width=7cm,height=5cm]{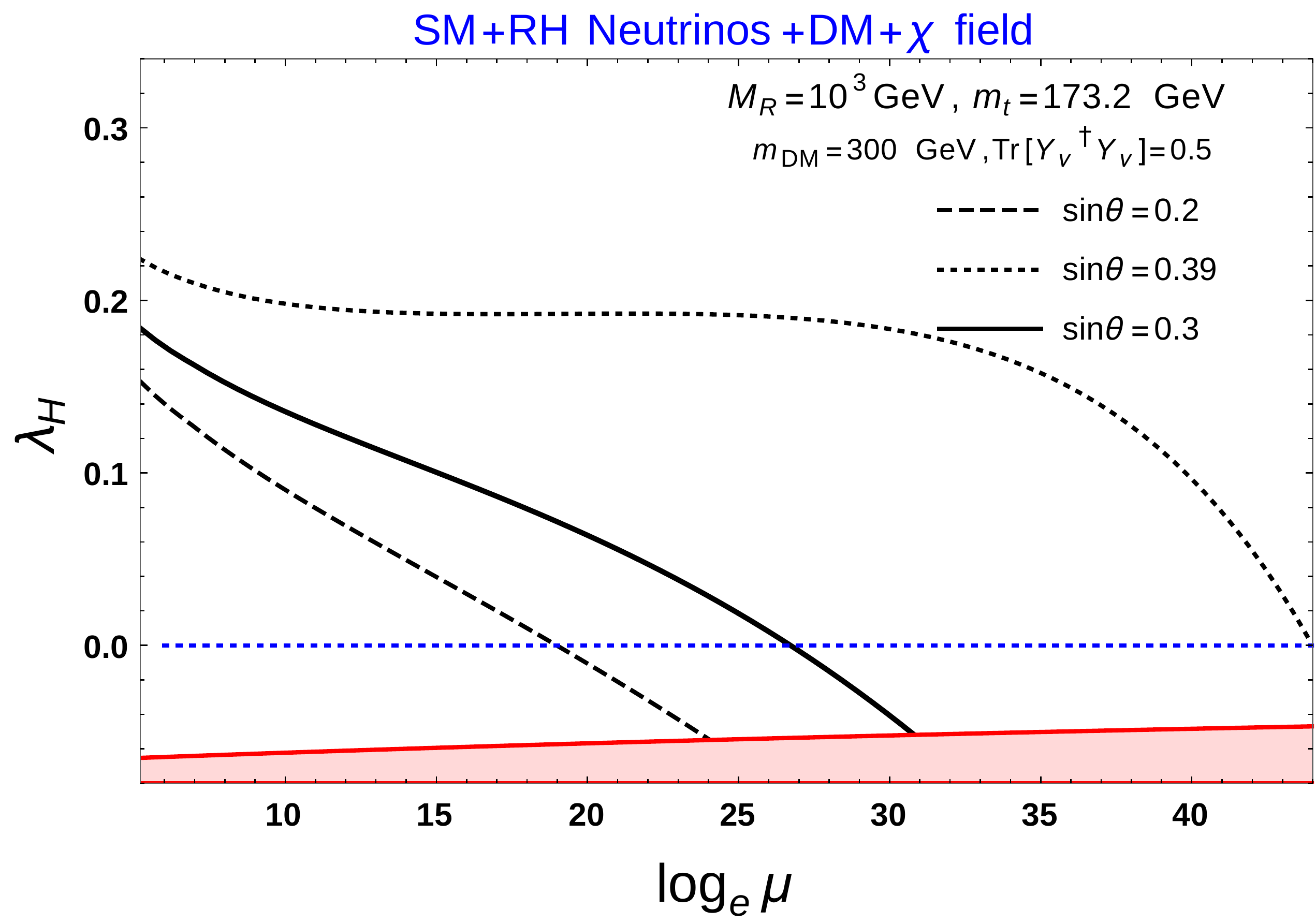}
 \end{center}
\caption{RG running of $\lambda_H$ with energy scale $\mu$ for different values of $\sin\theta$
 in the combined set up of SM+DM+RH neutrinos+$\chi$ field where in [left panel] $M_R=10^8$ GeV and in
 [right panel] $M_R=10^{3}$ GeV. Other reference values: $m_{\textrm{DM}}=300$ GeV, $m_{H_2}=300$ GeV, 
$\textrm{Tr}[Y_\nu^\dagger Y_\nu]=0.5$ and $\lambda_{\phi H}=0.06$ and $\lambda_{\chi\phi}=0.135$. }
  \label{fig:running4}
 \end{figure} 
we study the running of the Higgs quartic coupling $\lambda_H$ from EW scale to $M_P$ as shown in Fig.\ref{fig:running3}. 
In obtaining the running, we have considered $m_{H_2} = 300$ GeV, $\sin\theta =0.2$ and $m_{\textrm{DM}}$ is considered to be 
300 GeV. The values of $\lambda_{\chi \phi}$ and $\lambda_{\phi H}$ are fixed at 0.135 and 0.06  respectively (this particular 
point is denoted by a blue dot, named $\small{X}$, on Fig.\ref{fig:sinth} ). It turns out that any 
other set of $\lambda_{\chi \phi}$ and 
$\lambda_{\phi H}$ other than this blue dot from Fig. (while $m_{\textrm{DM}} = 300$ GeV is fixed) would not change our 
conclusion significantly as long as $\sin\theta$ is considered at 0.2. In order to compare the effect of the extra scalar $\chi$ 
in the theory, we keep the neutrino parameters Tr[$Y^{\dagger}_{\nu} Y_{\nu}$] and $M_R$ at their respective values 
considered in Figs.\ref{fig:SMrunning}, \ref{VSDMrunning}.

\begin{figure}[htb!]
 \begin{center}
 \includegraphics[width=8cm,height=6cm]{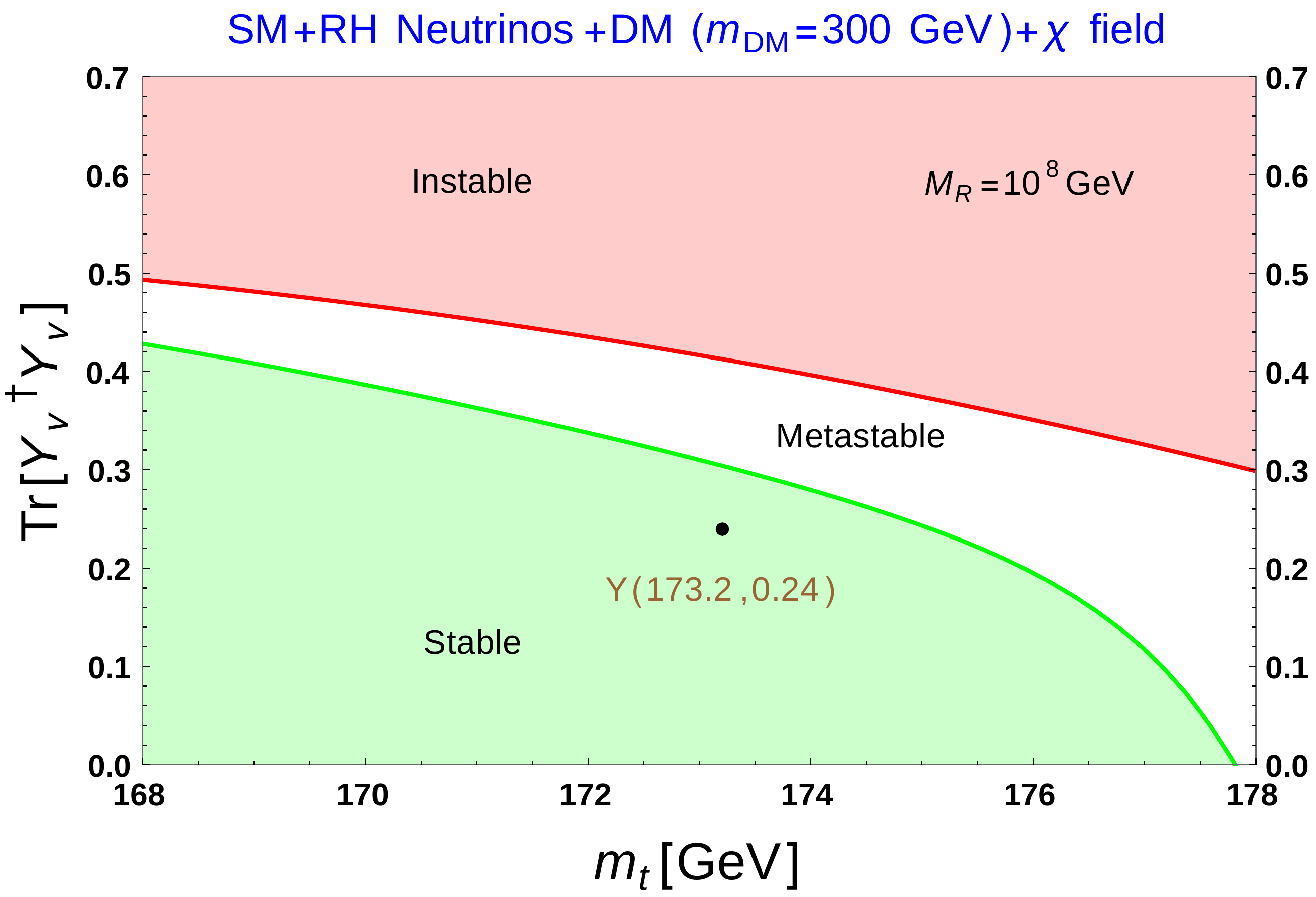}
 \end{center}
\caption{Stability, metastability and instability region on 
$~\textrm{Tr}[Y_{\nu}^\dagger Y_\nu]$ -$m_t$ plane for $M_R=10^8$ GeV in the
extended scenario of SM with 
3 RH neutrinos, DM and $\chi$. We have used point X ($\lambda_{\phi H}=0.06$, $\lambda_{\chi\phi}=0.135$) from 
Fig.\ref{fig:sinth}, $\sin\theta=0.2$, $m_{H_2}=300$ GeV, 
$v_\chi=800$ GeV, $m_{\textrm{DM}}=300$ GeV and $\lambda_{\phi}= 0.7$ as benchmark points. }
  \label{fig:SMDMEx_contMN108}
 \end{figure}

 \begin{figure}[htb!]
 \begin{center}
$$
\includegraphics[width=8cm,height=6cm]{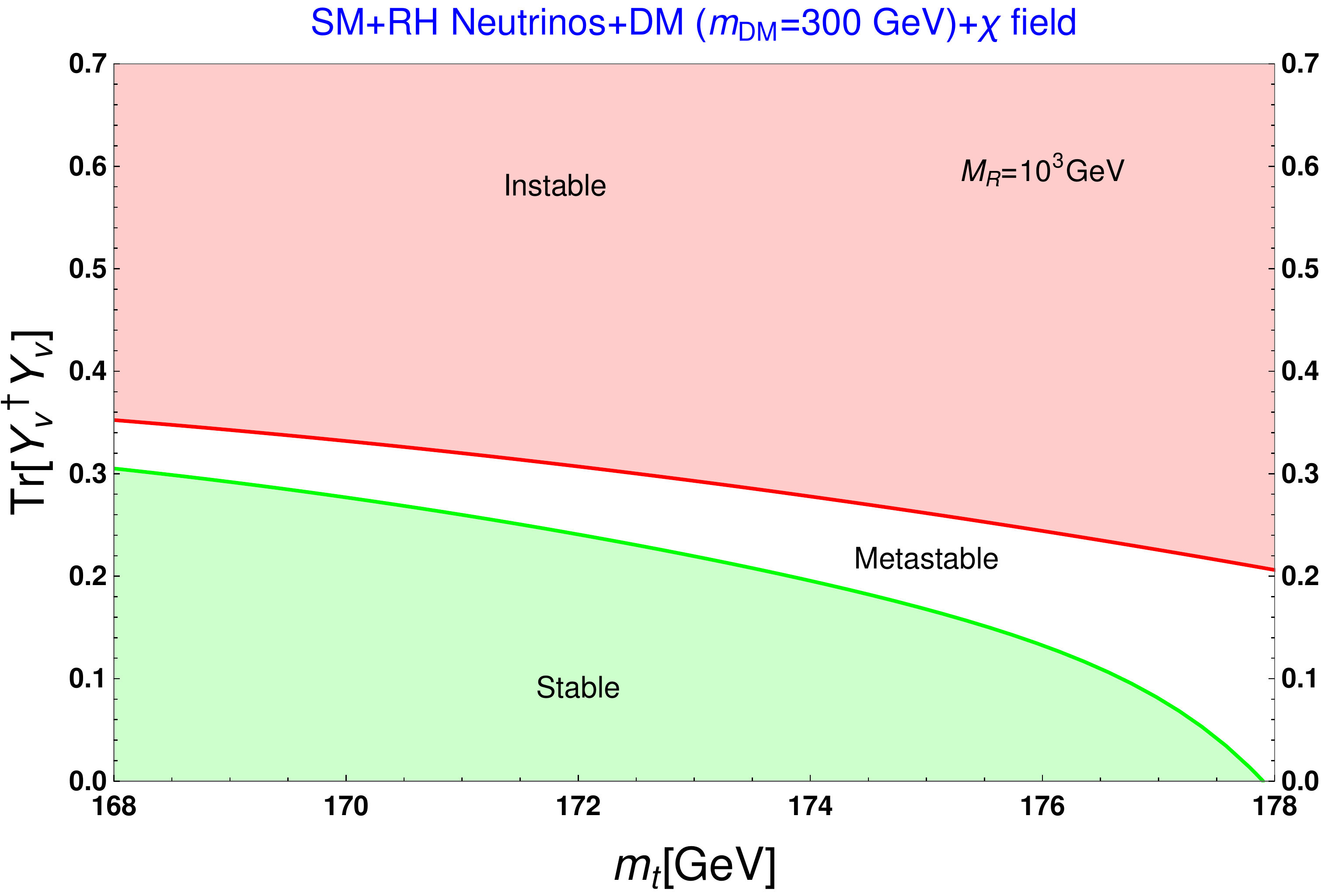}
\includegraphics[width=8cm,height=6cm]{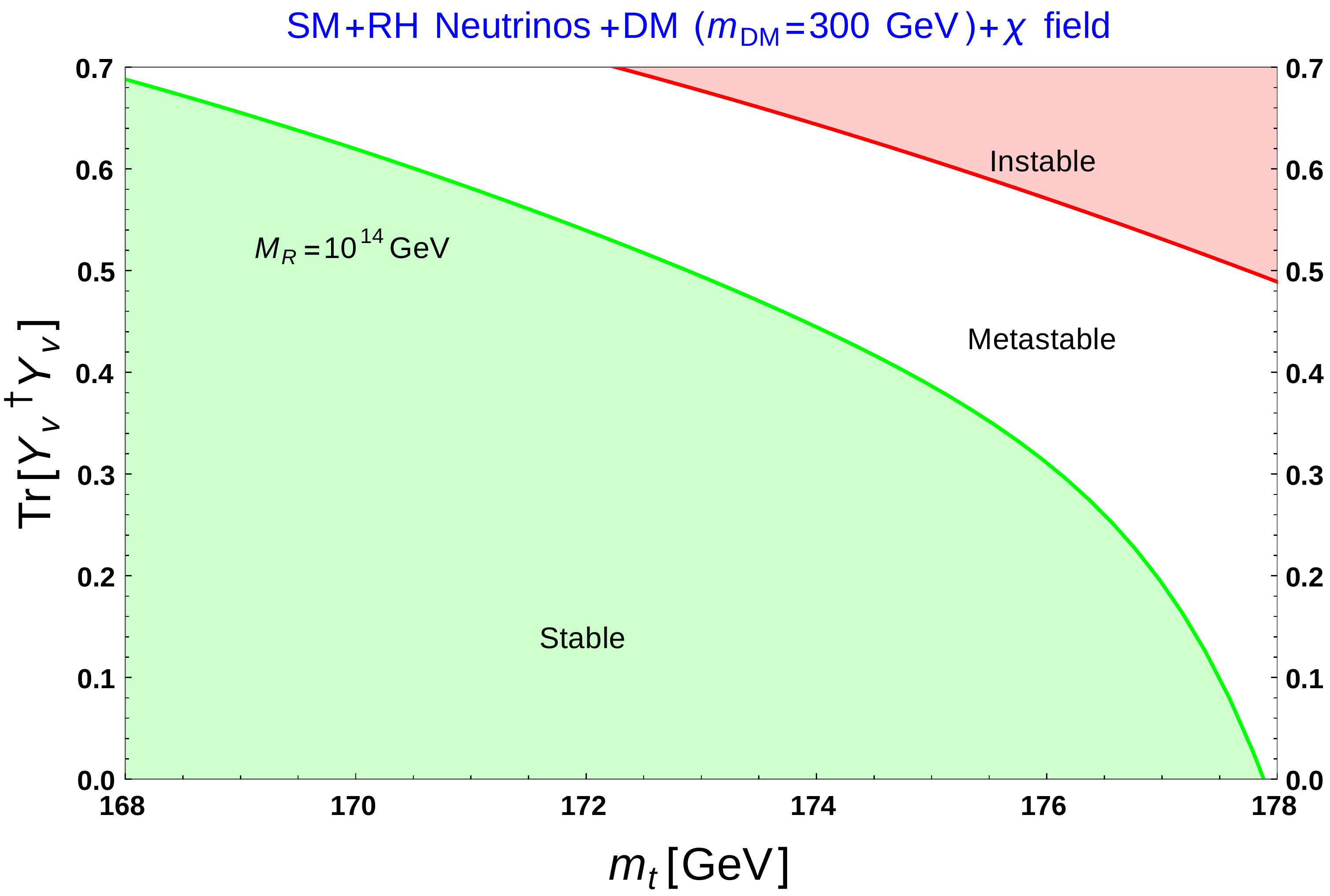}
$$
 \end{center}
\caption{Stability, metastability and instability region on 
$~\textrm{Tr}[Y_{\nu}^\dagger Y_\nu ]$ -$m_t$ plane in the
extended scenario of SM with 
3 RH neutrinos, DM and $\chi$ for (right panel) $M_R=10^3$ GeV and (left panel) $M_R=10^{14}$ GeV.
 We have used Point X ($\lambda_{\phi H}=0.06$, $\lambda_{\chi\phi}=0.135$) from 
Fig.\ref{fig:sinth}, $\sin\theta=0.2$, $m_{H_2}=300$ GeV, 
$v_\chi=800$ GeV, $m_{\textrm{DM}}=300$ GeV and $\lambda_{\phi}= 0.7$ as benchmark points. }
  \label{fig:SMDMEx_cont}
 \end{figure}

In the left panel of Fig.\ref{fig:running3}, the running is performed for three different choices of $M_R$, specifically at 
1 TeV, $10^8$ GeV and $10^{14}$ GeV while top mass is fixed at 173.2 GeV. A similar plot is exercised in right panel of Fig.\ref{fig:running3} where three different choices of $m_t = (171, 173.2, 177)$ GeV are considered while 
$M_R$ is fixed at $10^8$ GeV. 
Contrary to our previous finding in section  (see Fig.\ref{fig:SMrunning},
 \ref{VSDMrunning} ), we clearly see here that with $M_R=10^{14}$ GeV and $m_t=171$ GeV, $\lambda_{H}$ remains 
positive upto $M_P$ even in presence of large  Tr[$Y^{\dagger}_{\nu} Y_{\nu}$]  $\sim \mathcal{O}(1)$. Hence EW 
vacuum turns out to be absolutely stable. Although there exists other values of $M_R$ and/or $m_t$, for which  
EW vacuum still remains unstable, the scale at which $\lambda_H$ enters into the instable region is getting delayed 
with a noticeable change from earlier cases (Figs.\ref{fig:SMrunning}, \ref{VSDMrunning}). This becomes possible 
due to the introduction of the $\chi$ field having contribution mostly from the $\sin\theta$ parameter. 
In order to show its impact on stability, in Fig.\ref{fig:running4} (left panel), we plot $\lambda_{H}$ running with 
different choices of $\sin\theta = 0.1, 0.2, 0.3$ for $M_R=10^8$ GeV, $m_t=173.2$ GeV and $m_{\textrm{DM}}=300$ GeV 
while keeping $\textrm{Tr}[Y_\nu Y_\nu^\dagger]=0.5$ (same as in Fig.\ref{fig:running3}, left panel, black solid line). 
It shows that while $\sin\theta = 0.2$ (black solid line) can not make the EW vacuum absolutely stable till $M_{P}$, 
an increase of $\sin\theta$ value $\sim$ 0.3 can do it (dotted line). Similarly in Fig. \ref{fig:running4} (right panel), 
we consider a lowerer $M_R$ as 1 TeV. We have already noticed that such a low $M_R$ with large $\textrm{Tr} 
[Y_\nu^\dagger Y_\nu]=0.5$ pushes EW vacuum toward instability at a much lower scale $\sim 10^6$ GeV. In order 
to make the EW vacuum stable with such an $M_R$ and $\textrm{Tr}[Y_\nu^\dagger Y_\nu]$, one requires 
$\sin\theta \sim 0.4$ as seen from the right panel of Fig.\ref{fig:running4} (dotted line). However such a large 
$\sin\theta$ is ruled out from the experimental constraints \cite{Robens:2016xkb}. For representative purpose, 
we also include study with other $\sin\theta = 0.2, 0.3$ denoted by dashed and solid lines.

\begin{figure}[htb!]
 \begin{center}
\includegraphics[width=8cm,height=6cm]{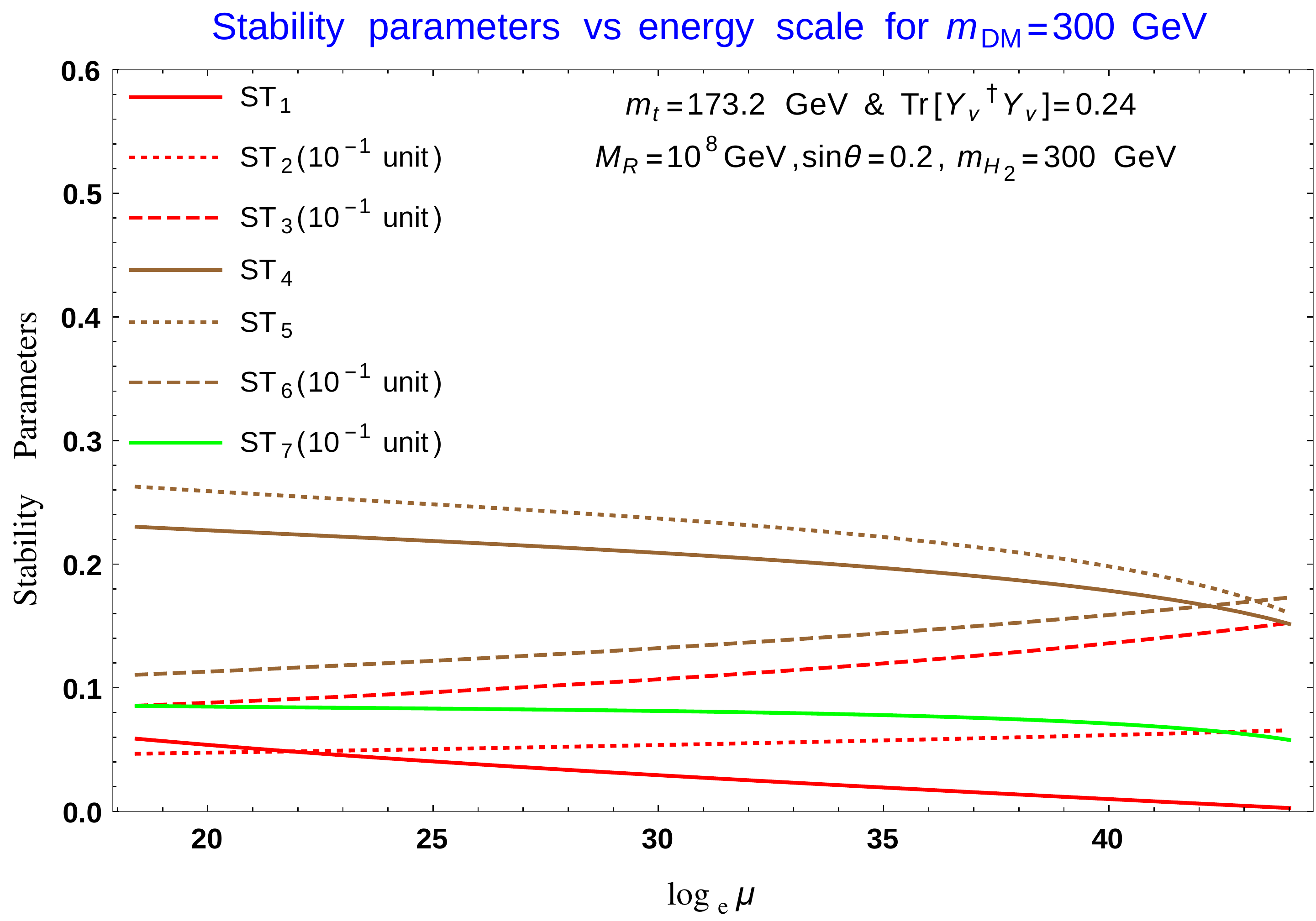}
 \end{center}
\caption{Evolution of stability parameters (Eq.\ref{stabC}) for the point Y ($m_t=173.2$ GeV, Tr$[Y_\nu^\dagger Y_\nu]=0.24$)
 from Fig.\ref{fig:SMDMEx_cont} (top right panel).
 Benchmark points: Point X ($\lambda_{\phi H}=0.06$, $\lambda_{\chi\phi}=0.135$) from Fig.\ref{fig:sinth}, $M_R=10^8$ GeV, $\sin\theta=0.2$, $m_{H_2}=300$ GeV, 
$v_\chi=800$ GeV, $m_{\textrm{DM}}=300$ GeV and $\lambda_{\phi}= 0.7$ have been used. }
  \label{fig:StabPara}
 \end{figure}

\begin{figure}[htb!]
 \begin{center}
\includegraphics[width=8cm,height=6cm]{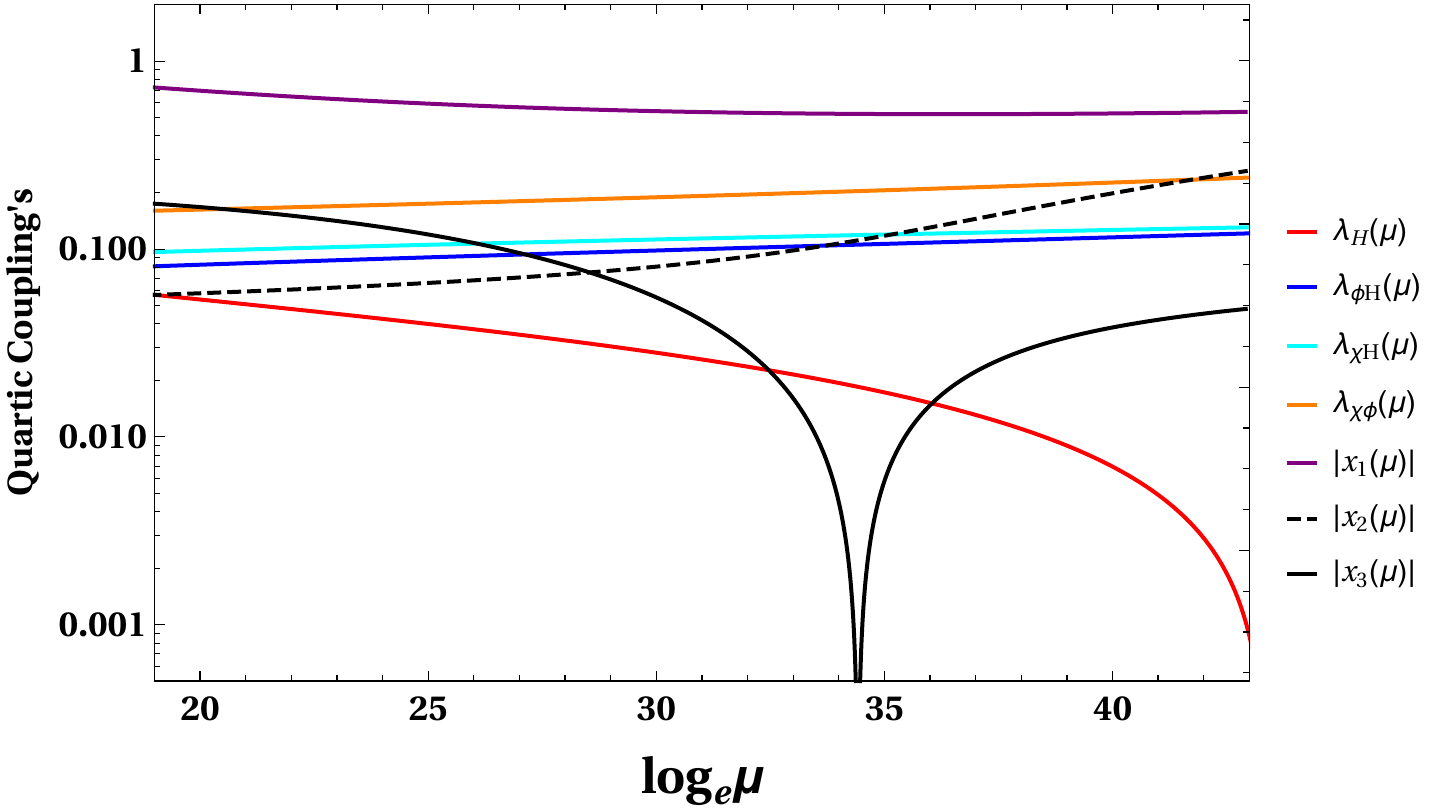}
 \end{center}
\caption{Evolution of parameters required to satisfy the perturbativity unitarity limit (Eq.(\ref{puc}))
 for the point Y ($m_t=173.2$ GeV, Tr$[Y_\nu^\dagger Y_\nu]=0.24$)
 from Fig.\ref{fig:SMDMEx_cont} (top right panel).
 Benchmark points: Point X ($\lambda_{\phi H}=0.06$, $\lambda_{\chi\phi}=0.135$) from Fig.\ref{fig:sinth}, $M_R=10^8$ GeV, $\sin\theta=0.2$, $m_{H_2}=300$ GeV, 
$v_\chi=800$ GeV, $m_{\textrm{DM}}=300$ GeV and $\lambda_{\phi}= 0.7$ have been considered. }
  \label{fig:pu}
 \end{figure}

We provide Fig.\ref{fig:SMDMEx_contMN108} where the regions with stability, meta-stability and instability are 
marked green, white and pink patches in the plane containing Tr[$Y^{\dagger}_{\nu} Y_{\nu}$] and $m_t$. 
With $M_R=10^3$ GeV and $M_R=10^{14}$ GeV, similar plots are shown in Fig.\ref{fig:SMDMEx_cont},  
left and right panels. Finally in Fig.\ref{fig:StabPara}, we have shown the RG evolution of all the stability conditions in Eq.(\ref{stabC}) 
from $m_t$ to $M_P$ to check their validity all the way upto $M_P$.  For this purpose, we have considered the 
initial values of the parameters involved in the following way. For values of $\lambda_{\phi H}$ and 
$\lambda_{\chi \phi}$ corresponding to $\sin\theta = 0.2, v_{\chi} = 800$ GeV and $m_{\textrm{DM}} = 300$ GeV, 
we have considered the benchmark point values as indicated by a blue dot named $X$ in Fig.\ref{fig:sinth} . 
The value of $\lambda_{\chi}$ is then followed from Eq.(\ref{lambdaChi}) and $\lambda_{\phi}$ is chosen to be at 0.7.  
Values of Tr[$Y^{\dagger}_{\nu} Y_{\nu}] = 0.24$ and $m_t = 173.2$ GeV are chosen for this purpose from 
Fig.\ref{fig:SMDMEx_contMN108} (here the benchmark values are denoted by a black dot $Y$). We conclude that all the stability criteria 
are fulfilled within the framework. Lastly we comment that instead of picking up the point X from relic density 
contour with $\sin\theta\sim 0.2$ in Fig.\ref{fig:sinth} to study vacuum stability in our model, we could have 
chosen any other point from that curve. As the stability of Higgs vacuum primarily depends on the value 
of $\theta$, our conclusion would not change much. However choice of any point having large $\lambda_{\chi\phi}$ 
could make it reaching Landau pole well before $M_P$ in its RG running through Eq.(\ref{RGlambdachiphi}).
To avoid that one can reduce the value of $\lambda_{\phi}$ $\sim\mathcal{O}(10^{-2})$ or less (earlier it was 
0.7) which has no direct connection or impact on DM phenomenology and vacuum stability analysis in the proposed set up. In Fig.\ref{fig:pu}, we have
shown the running of all parameters from $M_R$ to $M_P$ involved in perturbative unitarity bound for the  benchmark point: $m_{H_2}=300$ GeV, $\tan\beta=0.30$,
 $\sin\theta=0.2$, $m_{\textrm{DM}}=300$ GeV, $\lambda_{\phi H}$=0.06, $\lambda_{\chi\phi}
=0.135$, $M_R=10^8$ GeV and Tr$[Y_\nu^\dagger Y_\nu]=0.24$ with $m_t=173.2$ GeV. The parameters never exceed the upper limits coming from
the unitarity bound.  We have also confirmed that any other benchmark points wherever mentioned
in our analysis satisfy the perturbativity unitarity limit.

We end this section by comparing the results of vacuum stability in presence of 
(i) only RH neutrinos, (ii) RH neutrinos + DM and (iii) RH neutrinos + DM + extra scalar with non-zero vev, where
in each cases neutrino Yukawa coupling $Y_\nu$ has sizeable contributions. For this purpose, we consider $m_t=173.2$ GeV and
 $M_R=10^8$ GeV. From Fig.\ref{fig:SMCont}, for SM + RH
 neutrinos, we see that stability can not be achieved.
 The metastability scenario is still valid in this case upto Tr$[Y_\nu^\dagger Y_\nu]<0.26$. Next we add a singlet scalar
 DM candidate with nonzero Higgs portal coupling to SM with RH neutrinos. Fig.\ref{VSDMcontour} (left panel) shows, for 
 $m_{\textrm{DM}}=300$ GeV, stability of EW vacuum still remains elusive. On the other hand the metastability
 bound on Tr$[Y_\nu^\dagger Y_\nu]$ increases slightly from previous limit to 0.28. So DM with mass 300 GeV has mild impact
 on study of vacuum stability. Finally we add the extra scalar singlet with non zero vev to the 
 SM with RH neutrinos and scalar DM. We have fixed the heavier Higgs mass $m_{H_2}=300$ GeV and $\sin\theta=0.2$.
 Now in the combined set up of SM, scalar DM, scalar with non zero vev and RH neutrinos, the situation 
 changes drastically from previous case as seen in Fig.\ref{fig:SMDMEx_contMN108}.
  For the same top and RH neutrino masses, we can now achieve absolute stability upto
 Tr$[Y_\nu^\dagger Y_\nu]<0.3$ and the metastability bound on Tr$[Y_\nu^\dagger Y_\nu]$ further improved to 0.41.
Overall notable enhancement in the stability and metastability region has been observed in Tr$[Y_\nu^\dagger Y_\nu]-m_t$ plane
compared to the earlier cases.
 Hence, the numerical comparison clearly shows that the extra scalar having non zero mixing with SM Higgs effectively
 plays the leading role to get absolute vacuum stability in our model.

\section{Connection with other observables}\label{Nu}
 In this section, 
we first discuss in brief the constraints on the parameters of the model that may arise from lepton flavor violating (LFV) decays.
 The most stringent limit follows from $\mu\rightarrow e\gamma$ decay process. The 
branching ratio of such decay process in our set-up is given by\cite{Ilakovac:1994kj,Tommasini:1995ii,Dinh:2012bp}
\begin{align}\label{Br}
\textrm{Br}(\mu\rightarrow e\gamma)=\frac{3\alpha_e v^4}{16\pi M_R^4}|Y_{\nu_{ei}}^\dagger Y_{\nu_{i\mu}}|^2|f(x)|^2,
\end{align}
where $\alpha_e=\frac{e^2}{4\pi}$ is the fine sructure constant, $i$ runs from 1 to 3, $x=\frac{M_R^2}{m_W^2}$ and
\begin{align}
f(x)=\frac{x \left(2 x^3+3 x^2-6 x-6 x^2 \ln x+1\right)}{2 (1-x)^4}.
\end{align}
The current experimental limit on LFV branching ratio is\cite{Agashe:2014kda}
\begin{align}\label{BrBound}
\textrm{Br}(\mu\rightarrow e\gamma)<5.7\times 10^{-13}.
\end{align}

\begin{figure}[htb!]
 \begin{center}
 \includegraphics[width=10cm,height=7cm]{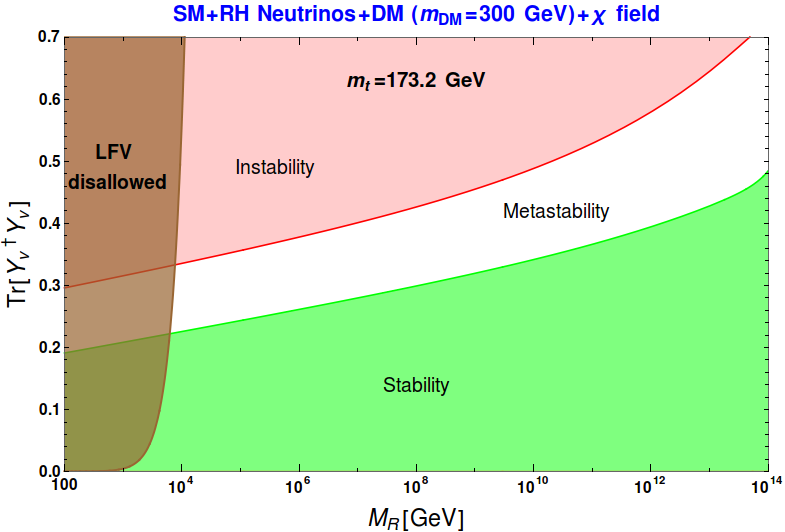}
 \end{center}
\caption{LFV and absolute vacuum stability constraint on  Tr$[Y_\nu^\dagger Y_\nu]-M_R$
 in the combined set up of SM+DM+RH neutrinos+$\chi$ field where  $m_{\textrm{DM}}=300$ GeV, $m_{H_2}=300$ GeV,
 $\sin\theta=0.2$, $\lambda_{\phi H}=0.06$ and $\lambda_{\chi\phi}=0.135$. }
  \label{fig:LFV}
 \end{figure}

Using this limit, we therefore obtain bounds on $|(Y_\nu^\dagger Y_\nu)_{e\mu}|$ corresponding to a fixed $M_R$ value
which can be converted to constrain $\textrm{Tr}[Y_\nu^\dagger Y_\nu]$ in our set up. In obtaining limits on $\textrm{Tr}[Y_\nu^\dagger Y_\nu]$ 
(for fixed $M_R$), first note that $Y_\nu^\dagger Y_\nu$ remains function of $M_R$ and parameter $a$ only (see Eq.(\ref{casas}) with $O=\mathbb{I}$),
once the best fit values of neutrino mixing angles\cite{Esteban:2016qun,deSalas:2017kay} are used to evaluate $U_{\textrm{PMNS}}$. Hence 
LFV limit basically constrains the parameter $a$ which in turn is used to obtain $\textrm{Tr}[Y_\nu^\dagger Y_\nu]$. This limit is
shown on Fig.\ref{fig:LFV} by the brown solid line, the left side of which is the disallowed region by LFV.

In the same plane of Fig.\ref{fig:LFV} we also include the region of the parameter space allowed by both stability and metastability criteria.
The green shaded region denotes the absolute stability of Higgs vacuum while the white region satisfies the metastability condition. 
 We also indicate the instability region by pink patch in the same figure under discussion. 
 For this purpose we have used $m_t=173.2$ GeV and $m_{\textrm{DM}}=300$ GeV, $\sin\theta = 0.2$,
  $\lambda_{\phi H} =0.06$ and $\lambda_{\chi\phi}=0.135$ (corresponding to the 
benchmark point indicated by X in Fig.\ref{fig:sinth}).
The brown shaded region is disfavored by the LFV constraint. Hence from Fig.\ref{fig:LFV} we infer that for low $M_R$,
LFV constraints turn out to be stronger one and for high $M_R$ values, $\textrm{Tr}[Y_\nu^\dagger Y_\nu]$ is mostly restricted by the stability issue.

It turns out that the proposed scenario does not provide any significant contribution to neutrinoless double beta 
decay\cite{Ibarra:2010xw,Tello:2010am,Mitra:2011qr,Chakrabortty:2012mh,Dev:2014xea,Dev:2013vxa}
even for relatively low RH neutrino mass ($\sim 10^3$ GeV). This is in line with the observation made in \cite{Bambhaniya:2016rbb}.
Before concluding the section, it is perhaps important to comment on the possibility of explaining the baryon asymmetry of the Universe (BAU).
The involvement of RH neutrinos would make the leptogenesis natural candidate to explain BAU
from the completion point of view. However
with the exactly degenerate RH neutrinos (we consider this for simplicity though), it is not possible. Once a small mass-splitting
$\Delta M_R$ between two heavy RH neutrinos can be introduced (for example by radiative 
effect \cite{GonzalezFelipe:2003fi,Turzynski:2004xy,Branco:2005ye}), resonant leptogenesis 
mechanism \cite{Flanz:1996fb,Pilaftsis:1997jf,Pilaftsis:2003gt} can be succesfully implemented\cite{Branco:2009by}.
 Apart from this, provided one can extend our vacuum stability
analysis in presence of non-degenerate RH neutrinos\cite{Coriano:2014mpa} with DM and $\chi$ 
field, usual thermal leptogenesis can also be employed to explain the
BAU of the universe.

\section{Conclusions}\label{Conclu}
We have considered an extension of the SM by three RH neutrinos and two scalar singlets with 
an aim to study the EW vacuum stability in a framework that can incorporate a stable light DM 
within the reach of collider experiments and to explain the light neutrino mass. A $Z_2 \times Z'_2$ symmetry is imposed of which $Z'_2$ is 
broken from the vev of one of the scalars. It is known that with a real scalar singlet DM model, 
present experimental limits by LUX 2016 and XENON 1T rule out DM mass below $m_{\textrm{DM}}=500$
GeV. Also its presence does not modify the fate of EW vacuum much and hence keep it metastable 
only. Although metastability is acceptable, it however leaves some unwanted questions if we include 
primordial inflation in the picture. So an absolute stability of the EW vacuum is more favourable. 
On the other hand, introduction of RH neutrinos would have large impact on the running of the 
Higgs quartic coupling due to the neutrino Yukawa interaction. Provided the neutrino Yukawa 
coupling is as large as $\mathcal{O}$(1) or more, it can actually destabilize the EW vacuum. Hence 
we have tried here achieving the stability of the EW vacuum in presence of RH neutrinos and 
DM. We also plan to find the possibility of a light scalar DM below 500 GeV. For 
this purpose, we have introduced additional scalar field which gets a vev. The other scalar among the 
two introduced does not get a vev and thereby is a good candidate for being a dark matter. The 
presence of the singlet with non-zero vev helps achieving the vacuum stability through a threshold 
like correction to $\lambda_H$.  So in this particular scenario {\it{i.e.}} SM extended by DM, three 
RH neutrinos plus one extra scalar, we have studied the Higgs vacuum stability issue considering 
large Yukawa coupling and variation of $m_t$ within $2\sigma$ range of uncertainty.
  We have found the 
stability region in the Tr$[Y_\nu^\dagger Y_\nu]-m_t$ plane has been significantly increased in presence of $\chi$.
Simultaneously mixing of this extra scalar with SM Higgs doublet ensures its involvement in the DM 
annihilations. This mixing is effectively controlled by the Higgs portal coupling of the scalar which 
also enters into the running of the Higgs quartic coupling. Hence an interplay between the two 
conditions: one is to achieve the EW vacuum stability and the other is to find a viable DM below 
500 GeV, can actually constrain the parameters involved to some extent. Since the set-up involves 
several new particles, finding their existence in future and ongoing experiments would be an interesting 
possibility to search for. Here we have assumed the physical Higgs other than the SM one is heavier. 
The other situation where the second physical Higgs is lighter than the Higgs discovered at 125 GeV. 
However this case is not of very interest in the present study as following from Eq.(\ref{lambdaH}), it can be seen that 
the effective Higgs quartic coupling becomes less than the SM one in this case and this would not help 
making EW vacuum stable. Also the $\sin\theta$ allowed region for $m_{H_2} < m_{H_1}/2$ is almost 
excluded from the decay of $H_2 \rightarrow H_1 H_1$. Hence we discard this possibility. One interesting 
extension of our work could be the study of a SM gauge extension where the involvement of gauges 
bosons can modify our result. We keep it for a future study. 


\appendix
\numberwithin{equation}{section}
\section*{Appendix}
\section{Unitarity Constraints}\label{appen1} 

In this section we draw the perturbative unitarity limits on quartic couplings present in our model. 
Scattering amplitude for any  $2 \to 2$ process can be expressed in terms of Legendre polynomial as\cite{Horejsi:2005da,Bhattacharyya:2015nca} 
$$\mathcal{M}^{2\rightarrow 2}=16 \pi \sum^{\infty}_{l=0}{a_l} (2l+1) \mathit{P}_l (\cos \theta) ,$$ 
where, $\theta$ is the scattering angle and $P_l(\cos\theta)$ is the Legendre polynomial of order $l$. 
 In high energy limit only s wave ($l=0$) partial amplitude $a_0$
 will determine the leading energy dependence of the scattering processes\cite{Horejsi:2005da,Bhattacharyya:2015nca}. The unitarity constraint
says
\begin{align}
 \lvert\rm Re~a_0 \rvert<1/2\label{eq:uni1}.
\end{align}
This constraint Eq.(\ref{eq:uni1}) can be further translated to a bound on the scattering amplitude $\mathcal{M}$\cite{Horejsi:2005da,Bhattacharyya:2015nca}.
\begin{align}
\lvert \mathcal{M} \rvert < 8 \pi.\label{eq:unitariC}
\end{align}
In our proposed model we have multiple possible $2\rightarrow 2$ scattering processes. Therefore we need to construct a matrix 
($M^{2\rightarrow2}_{i,j}=\mathcal{M}_{i\rightarrow j}$)
 considering all possible
two particle states.  
Finally we need to calculate the eigenvalues of $M$ and employ the bound as in Eq.(\ref{eq:unitariC}).

In the high energy limit we express the SM Higgs doublet as $H^T=(w^+, \frac{H^0+i z}{2})$. Then  
the scalar potential ($V$) in Eq.(\ref{eq:lag_tot}) gives rise to eleven neutral combination of two particle states
\begin{align}
w^+w^-,~  \frac{zz}{\sqrt{2}},~ \frac{H^0 H^0}{\sqrt{2}}, ~ \frac{\chi \chi}{\sqrt{2}}, ~\frac{\phi ~\phi}{\sqrt{2}},
 ~H^0 \chi,~ H^0~\phi,~ \chi~\phi,~z ~H^0, ~z ~\chi,~ z~\phi),
\end{align}
 and four singly charged two particle states
\begin{align}
 w^+H^0,~w^+\chi,~  w^+z,~ w^+\phi.
\end{align}
\noindent Hence we can write the scattering amplitude matrix ($M$) in block diagonal form by decomposing it 
into neutral and singly charged sector as
 \begin{eqnarray}
\label{eq:MT}
 M_{15\times 15}=\left(
\begin{array}{cc}
M^{\textrm{n}}_{11\times11} & 0\\
0 & M^{sc}_{4\times4}
\end{array}
\right) .
\end{eqnarray}
The submatrices are provided below :
\begin{eqnarray}
\label{eq:nc2}
 M^{n}_{11\times 11}=\left(
\begin{array}{ccccccccccc}
 4 \lambda_H & \sqrt{2} \lambda_H & \sqrt{2} \lambda_H & \frac{\lambda_{\chi H}}{\sqrt{2}} &
 \frac{\lambda_{\phi H}}{\sqrt{2}} & 0 & 0 & 0 & 0 & 0 & 0 \\
 \sqrt{2} \lambda_H & 3 \lambda_H & \lambda_H & \frac{\lambda_{\chi H}}{2} & 
\frac{\lambda_{\phi H}}{2} & 0 & 0 & 0 & 0 & 0 & 0 \\
 \sqrt{2} \lambda_H & \lambda_H & 3 \lambda_H & \frac{\lambda_{\chi H}}{2} &
 \frac{\lambda_{\phi H}}{2} & 0 & 0 & 0 & 0 & 0 & 0 \\
 \frac{\lambda_{\chi H}}{\sqrt{2}} & \frac{\lambda_{\chi H}}{2} & 
\frac{\lambda_{\chi H}}{2} & \frac{\lambda_{ \chi} }{2} & \frac{\lambda_{\chi \phi} }{2} & 0 & 0 & 0 & 0 & 0 & 0 \\
 \frac{\lambda_{\phi H}}{\sqrt{2}} & \frac{\lambda_{\phi H}}{2} &
 \frac{\lambda_{\phi H}}{2} & \frac{\lambda_{\chi \phi} }{2} & \frac{\lambda_{ \phi} }{2} & 0 & 0 & 0 & 0 & 0 & 0 \\
 0 & 0 & 0 & 0 & 0 & \lambda_{\chi H} & 0 & 0 & 0 & 0 & 0 \\
 0 & 0 & 0 & 0 & 0 & 0 & \lambda_{\phi H} & 0 & 0 & 0 & 0 \\
 0 & 0 & 0 & 0 & 0 & 0 & 0 & \lambda_{\chi \phi}  & 0 & 0 & 0 \\
 0 & 0 & 0 & 0 & 0 & 0 & 0 & 0 & 2 \lambda_H & 0 & 0 \\
 0 & 0 & 0 & 0 & 0 & 0 & 0 & 0 & 0 & \lambda_{\chi H} & 0 \\
 0 & 0 & 0 & 0 & 0 & 0 & 0 & 0 & 0 & 0 & \lambda_{\phi H} \\
\end{array}
\right) .
\end{eqnarray}

\begin{equation}
\label{eq:sc2}
M^{sc}_{4\times 4}=
\left(
\begin{array}{cccc}
 2\lambda_{ H} & 0 & 0 & 0 \\
 0 & \lambda_{ \chi H} & 0 & 0 \\
 0 & 0 & 2\lambda_{ H} & 0 \\
 0 & 0 & 0 & \lambda_{ \phi H} \\
\end{array}
\right) ~~.
\end{equation}
\noindent The distinct eigen values of matrix Eq.(\ref{eq:nc2}) and Eq.(\ref{eq:sc2}) are following :
$$ 2 \lambda_H,~\lambda_{\chi H}, ~\lambda_{\phi H},~\lambda_{\chi \phi} \textrm{ and } x_{1,2,3}, $$
where $x_{1,2,3}$ are the roots of the following polynomial equation, 
\begin{eqnarray}\label{eq:polroot}
x^3+x^2(-12 \lambda_{ H}-\lambda \chi -\lambda \phi )+x\left(12 \lambda_{ H}\lambda_{ \chi} +12 \lambda_{ H}\lambda_
{ \phi} -4 \lambda_{\chi H}^2-\lambda_{\chi_\phi} ^2+\lambda _\chi \lambda_{ \phi} -4 \lambda_{ \phi H}^2\right)\nonumber \\
 +12 \lambda_{ H}\lambda_{ \chi\phi}^2 -12 \lambda_{ H}\lambda_{ \chi} \lambda_{ \phi} +4 \lambda_{\chi H}^2 \lambda_{ \phi}
 + 4\lambda_{ \chi}  \lambda_{ \phi H}^2-8 \lambda_{\chi H}\lambda_{ \chi \phi}  \lambda_{ \phi H}= 0 .
\end{eqnarray}
Therefore the unitarity constraints in the proposed set up are following: 
\begin{eqnarray}
 \lambda_{H} ~ < 4 \pi ,~~~~  \lambda_{\phi H} ~ < 8 \pi ,~~~~  \lambda_{\chi H} ~ < 8 \pi ,~~~~  \lambda_{\chi \phi} ~ < 8 \pi ~~~ 
{\rm and} ~~ x_{1,2,3} < 16 \pi    ~~.
\label{roots}
 \end{eqnarray}


\end{document}